\documentclass[10pt, journal]{IEEEtran}

\usepackage{epsfig,color,amsmath,cite}
\usepackage{amsthm} %defined already in ieeeconf
\usepackage{amsmath}    %For theorems
\IEEEoverridecommandlockouts
\usepackage{empheq}
\usepackage{bm}
\usepackage{epstopdf}
\usepackage{amssymb}
\usepackage{url}
\usepackage{enumitem} %defined already in ieeeconf
\usepackage{multirow}
\usepackage{hhline}
\usepackage{booktabs}
\usepackage{mathtools}
\usepackage{makecell}
\usepackage[linesnumbered,boxed,commentsnumbered,ruled,vlined,longend]{algorithm2e}
\usepackage{comment}

\DeclareMathOperator{\rank}{rank}

\DeclareMathOperator*{\subjectto}{subject\ to}
\DeclareMathOperator*{\sto}{s.to}

\DeclareMathOperator{\Blkdiag}{Blkdiag}
\makeatother
\DeclareMathAlphabet\mathbfcal{OMS}{cmsy}{b}{n}

\newtheorem{theorem}{Theorem}

\newtheorem{mylem}{Lemma}

\newtheorem{myrem}{Remark}
\newtheorem{asmp}{Assumption}



\usepackage{stackengine}

\newcommand{\mat}[1]{\boldsymbol{#1}}

\newcommand{\bmat}[1]{\begin{bmatrix} #1 \end{bmatrix}}

\providecommand{\mA}{\ensuremath{\mat{A}}}

\newcommand{\m}{\boldsymbol}
\allowdisplaybreaks[4]
\usepackage[colorlinks = true,
linkcolor = blue,
urlcolor  = blue,
citecolor = blue,
anchorcolor = blue]{hyperref}

\newcommand{\mc}[1]{\mathcal{#1}}
\newcommand{\mbb}[1]{\mathbb{#1}}
\newcommand{\mr}[1]{\mathrm{#1}}
\usepackage[framemethod=TikZ]{mdframed}
\mdfdefinestyle{MyFrame}{%
	linecolor=black,
	outerlinewidth=1.25pt,
	roundcorner=1.25pt,
	innerrightmargin=5pt,
	innerleftmargin=5pt,}
	
\usepackage[noabbrev]{cleveref}

\usepackage{mathtools}

\DeclarePairedDelimiter\abs{\lvert}{\rvert}%
\DeclarePairedDelimiter\norm{\lVert}{\rVert}%

\makeatletter
\let\oldabs\abs
\def\abs{\@ifstar{\oldabs}{\oldabs*}}
\let\oldnorm\norm
\def\norm{\@ifstar{\oldnorm}{\oldnorm*}}
\makeatother

\usepackage[english]{babel}
\usepackage[utf8]{inputenc}
\usepackage[super]{nth}

\usepackage{graphicx}
\usepackage{float}
\usepackage[caption = false]{subfig}

\usepackage{array}
\usepackage{threeparttable}

\usepackage[english]{babel}
\usepackage[utf8]{inputenc}
\usepackage[super]{nth}

\RequirePackage{filecontents}

\def\newqed{{\null\nobreak\hfill\color{black}\ensuremath{\blacksquare}}}

\SetKwRepeat{Do}{do}{while}%

\newcommand*\widefbox[1]{\fbox{\hspace{0em}#1\hspace{0em}}}

\title{\centering \Huge {{ Observers for Differential Algebraic Equation Models of Power Networks: Jointly Estimating Dynamic and Algebraic States}}}

\author{Sebastian A. Nugroh$\text{o}^{\star}$, Ahmad F. Tah$\text{a}^{\dagger}$, Nikolaos Gatsi$\text{s}^{\ddagger}$, and Junbo Zha$\text{o}^{\diamond}$
	\thanks{
		$^{\star}$Department of Electrical Engineering and Computer Science, University of Michigan, 1301 Beal Ave., Ann Arbor, MI 48109 (snugroho@umich.edu).
		
		$^\dagger$Department of Civil and Environmental  Engineering, Vanderbilt University,  2201 West End Ave., Nashville, TN 37235 (ahmad.taha@vanderbilt.edu).
		
		$^{\ddagger}$Department of Electrical and Computer Engineering, The University of Texas at San Antonio, 1 UTSA Circle, San Antonio, TX 78249 (nikolaos.gatsis@utsa.edu).	
		
		$^{\diamond}$Department of Electrical and Computer Engineering, University of Connecticut, Storrs, CT 06269 (junbo@uconn.edu).
		
		$^\dagger$Corresponding author. 
		
	This work is partially supported by The Valero Energy Corporation and National Science Foundation (NSF) under Grants 1719043, 1847125, 2151571, and 2152450.}
}
 
\begin{document}

\newdimen\origiwspc%
\newdimen\origiwstr%
\origiwspc=\fontdimen2\font% original inter word space
\origiwstr=\fontdimen3\font% original inter word stretch

\fontdimen2\font=0.63ex% inter word space
%\fontdimen3\font=\origiwstr% inter word stretch

\maketitle
\thispagestyle{plain}
\pagestyle{plain}

\begin{abstract}
Phasor measurement units ({PMUs}) have become instrumental in modern power systems for enabling real-time, wide-area monitoring and control. Accordingly, many studies have investigated efficient and robust dynamic state estimation (DSE) methods in order to accurately compute the dynamic states of generation units. Nonetheless, most of them forego the dynamic-algebraic nature of power networks and only consider their nonlinear dynamic representations. Motivated by the lack of DSE methods based on power network's differential-algebraic equations (DAEs), this paper develops a novel observer-based DSE framework in order to perform simultaneous estimation of the dynamic and algebraic states of multi-machine power networks. Specifically, we leverage the DAE dynamics of a power network around an operating point and combine them with a PMU-based measurement model  capable of capturing bus voltages and line currents. The proposed $\mathcal{H}_{\infty}$ observer, which only requires detectability and impulse observability conditions which are satisfied for various power networks, is designed to handle various noise, unknown inputs, and input sensor failures.
The results obtained from performing extensive numerical simulations on the IEEE $9$-bus and $39$-bus systems showcase the effectiveness of the proposed approach for DSE purposes.
\end{abstract}

\begin{IEEEkeywords}
Transmission power networks, dynamic state estimation, differential-algebraic equation, robust $\mathcal{H}_{\infty}$ observer.  
\end{IEEEkeywords}

\vspace{-0.3cm}
\section{Introduction}

\IEEEPARstart{W}{ith} the vast utilization of Phasor Measurement Units (PMUs), power systems monitoring can be conveniently performed in real-time, thereby paving the possibility of implementing estimation-based feedback controls on power networks \cite{Chakrabortty2013}. Consequently, this results in the prevalence of research on dynamic state estimation (DSE) in the past two decades. Numerous approaches have been proposed accordingly. {For instance, Kalman Filter (KF)-based DSE and its variants, such as Extended Kalman Filter, Unscented Kalman Filter, Ensemble Kalman Filter, as well Particle Filter-based approach have been investigated for centralized/decentralized DSE \cite{Zhao2019PSDSE}.} In addition to these stochastic approaches, deterministic observers and their applications in DSE have also been recently proposed and studied. For example, the work in \cite{Qi2018Access} presents a comparative study between KF-based methods with a state observer designed for nonlinear systems, while \cite{Nugroho2020DSE} focuses on the estimation of a synchronous generator's internal states based on a nonlinear ordinary differential equation (ODE) model. 

In spite of these advances, the majority of literature on DSE overlook the dynamic-algebraic structure of power networks in their investigations, especially in a centralized DSE framework for multi-machine power networks. { For example, the works in \cite{Zhao2017EKF,Ghahremani2016,Qi2018Access,Qi2018UKF} consider nonlinear differential representations of power networks---usually obtained by means of the Kron reduction technique---which only consist of generators' dynamics. Although the resulting ODE models have a smaller number of states, the elimination of algebraic constraints and variables introduces several downsides.} First of all, it is argued in \cite{Gross2016} that the nonlinear ODE model may fail to accurately capture the power network's dynamics with respect to topological changes, which can be triggered by faults. {Secondly, it is not clear how  more detailed loads' dynamics can be incorporated on the basis of ODE models. Thirdly, if nonlinear ODEs of power networks are employed, it is difficult to study the DSE in which the PMUs are also installed on non-generator buses (such as load buses and non-generator unit buses) as the relations between bus voltages at non-generator buses and generators' states need to be established.} 

{In addition to the above arguments, PMUs are initially designed to measure voltage and current phasors in power systems \cite{sauer2017power} and some literature focusing on the state estimation problem in power systems are considering that a PMU installed on a bus is able to measure the complex quantities of bus voltage and line currents \cite{gomez2011use,Risbud2016,Sarri2016,Risbud2020}. Hence, the use of such measurement model in DSE analysis should have been widely considered. }
{Indeed, there exist approaches for DSE based on such measurement model. One of the prominent approaches is referred to as the \textit{two-stage} method \cite{Rouhani2018,Abur2015,Zhang2014}. In the first stage, the algebraic variables (e.g. bus voltages) are estimated based on information collected from PMUs and in the second stage, the estimated bus voltages are used by the state estimator to estimate the actual generators' variables. For instance, the works in \cite{Rouhani2018,Abur2015} implement the \textit{Least Absolute Value} (LAV) method in the first stage and Unscented KF in the second stage, whereas in \cite{Zhang2014}, the authors develop an adaptive KF to deal with unknown dynamics and measurement errors in the first stage while utilizing the Extended KF to estimate the dynamic states. It is important to note that, albeit this two-stage approach retains the dynamic-algebraic structure of the power network, two state estimation algorithms have to be performed separately, thereby increasing the computational burden required for performing DSE.}

{Motivated by the aforementioned limitations, in this work, we exploit the simultaneous dynamic \textit{and} algebraic nature of power networks to perform DSE from a descriptor system and control-theoretic perspective. Indeed, this particular research direction is less prominent in power systems literature. 
One of the earliest works in this direction is found in \cite{Dafis2002}, where the authors study the observability of power networks based on its DAE model using a linearization technique. 
Recently, a DSE approach in which distributed sliding-mode and algebraic observers are employed simultaneously to estimate each generator's electrical frequency is proposed in \cite{Rinaldi2017a} whereas in \cite{Rinaldi2017b}, high-order sliding mode observers are developed to estimate generators' rotor speeds and complex bus voltages based on simple nonlinear, power network differential-algebraic equations (DAEs)---both works become the basis of \cite{Rinaldi2020}. These studies, however, are conducted based on an overly simplified power network model using some assumptions including the use of a \nth{2}-order swing equation to represent generators' dynamics, the negligence of transmission lines' resistances and reactive power flows, the consideration of a uniformly-flat bus voltage profile, and PMUs measuring generator's rotor angle installed only on generator buses.
Such simplification may not be sufficient to aid frequency regulation in power systems through an output-feedback control framework---for instance, to perform load following control \cite{Taha2019TCNS}. 
A new approach to estimate both dynamic and algebraic states of generators in a decentralized framework is recently proposed in \cite{LorenzMeyer2020}, where an algebraic observer is developed to estimate the
load angle and quadrature-axis internal voltage of each generator. In order to estimate the relative rotor speed, the authors in \cite{LorenzMeyer2020} combine the Immersion
and Invariance technique \cite{astolfi_nonlinear_2008} with the  Dynamic Regressor and Mixing \cite{Aranovskiy2017} in their observer design. Due to the decentralized fashion however, this DSE approach relies on PMUs placed on generator terminals.  } 

{To that end, in this paper we construct a novel DSE framework built upon DAE representations of power networks which dynamics are studied around an operating point.  
This framework considers the aforementioned measurement model in which PMUs are capable of measuring bus voltage and line current phasors---such model allows PMUs to be placed arbitrarily: on generator, load, or even non-generator unit buses. Unlike other DSE methods, e.g. in \cite{Qi2018Access,Emami2015PF,Ghahremani2016,Qi2018UKF} and also others that are based on decentralized DSE framework such as \cite{LorenzMeyer2020}, this feature allows for flexible PMU placements since every generator terminal may not be equipped with a PMU \cite{Rouhani2018}. 
In contrast to \cite{Rinaldi2017a,Rinaldi2017b,Rinaldi2020}, our work herein utilizes a DAE representation of power networks with \textit{(i)} more detailed and comprehensive \nth{4}-order generator's transient dynamics, stator's algebraic constraints, generator's real and reactive power, and the network's complex power balance equations, and \textit{(ii)} more realistic and practical PMU-based measurement equations.} 

{To perform the DSE, a new robust $\mathcal{H}_{\infty}$ observer design for linear DAEs is introduced. Indeed, there exist several observer designs developed in the literature especially for linear DAEs and descriptor systems \cite{Darouach1995,Hou1999}. For example, to provide a more robust state estimates in the presence of disturbance, numerous $\mathcal{H}_{\infty}$ observers are proposed in \cite{darouach2009,Xu2003,XU200748}. 
Although the observers developed in \cite{darouach2009,Xu2003,XU200748} do consider disturbance, due to their relatively complex linear matrix inequality (LMI) formulations, their applicability and effectiveness to perform DSE in power networks have never been studied. 
Our  $\mathcal{H}_{\infty}$ observer, on the other hand,  has much simpler LMI formulations and this allows it to be employed to perform DSE with minimal computational efforts.
The main contributions of in this paper include:}
\vspace{-0.05cm}
{
\begin{itemize}[leftmargin=*]
	\item 
	The unification of an in-depth differential-algebraic model of  power networks---encompassing generators' transient models, stator algebraic constraints, generators' complex power, and power balance equations---together with a PMU-based measurement model to allow a more realistic approach for performing DSE in power networks. The DAEs capture the power network's dynamics around a certain operating point and become the foundation for the observer-based DSE. A careful analysis on the regularity, index characterization, impulsiveness, detectability, and impulse observability of the linearized power network's DAEs is also presented. 
\item A novel LMI-based $\mathcal{H}_{\infty}$ DAE observer is introduced---this $\mathcal{H}_{\infty}$ observer is then employed to perform robust DSE for the linearized dynamics of power networks. The key features of this observer are attributed to the more straightforward design as well as the simplicity of the associated LMI formulations used to compute the observer gain matrix. 
In order to tackle unknown inputs, a robust proportional-integral (PI) version of the $\mathcal{H}_{\infty}$ DAE observer---developed by augmenting the original DAE dynamics with an unknown inputs model---is also proposed. Since we utilize a static-gain observer, the complexity of performing DSE is considerably much lower than using the two-stage approach as in \cite{Rouhani2018}. One primary advantage of the robust $\mathcal{H}_{\infty}$ DAE observer---as opposed to KF-based filters---is that it requires no information regarding the statistical properties of the noise.
\item The effectiveness of the proposed approach 
to address DSE in power networks is demonstrated through the IEEE $9$-bus and $39$-bus systems, in which the proposed $\mathcal{H}_{\infty}$ DAE observer is tested for estimating generators' internal states and unmeasured bus voltages following a three-phase fault under various scenarios, including the presence of Gaussian and non-Gaussian process and measurement noise as well as unknown inputs. Numerical advantages of the developed observer over the two-stage approach \cite{Rouhani2018,Abur2015}, which combines LAV with KF in performing DSE are also showcased.
\end{itemize}  }
\vspace{-0.05cm}
A preliminary version of this paper was published in \cite{Nugroho2021CCTA}.
%The supplemental document \cite{Nugrohotcns2020} provides the complete proofs of the theorems presented in this paper.
The paper's organization is as follows. Section \ref{sec:modeling}, presents the linearized semi-explicit DAEs of power networks with a PMU-based measurement models while Section \ref{sec:reg_idx_obs} provides a brief study on the regularity, index characterization, and impulsiveness of the linear DAEs as well as some conditions to ensure detectability and impulse observability. In Section \ref{sec:DAE_observer}, simple Luenberger and robust $\mathcal{H}_{\infty}$ DAE observers are designed. Section \ref{sec:DAE_observer_UI} presents a way to modify the observers to handle unknown inputs and input sensor failures based on a PI framework. Thorough numerical studies are presented out in Section \ref{sec:numerical_tests} and the results are discussed accordingly. Finally, the paper is summarized in Section \ref{sec:conclusion}.% and possible future works are given.

\noindent \textit{\textbf{Notation.}} The notations $\mathbb{R}$ and $\mathbb{R}_{++}$ denote the set of real and positive real numbers, while the notation  $\mathbb{C}$ denotes the set of complex numbers. The notations $ {\mathbb{C}}_{-}$  and $\bar{\mathbb{C}}_{+}$ denote the set of all complex numbers such that the real parts are on the open left-half and closed right-half of the complex plane. The symbols $\mathbb{R}^n$ and $\mathbb{R}^{p\times q}$ denote the sets of row vectors with $n$ elements and matrices with size $p$-by-$q$ with elements in $\mathbb{R}$. 
The matrix $\m O$ denotes the zero matrix of appropriate dimension.
The operator $\mathrm{Blkdiag}(\cdot)$ constructs a block diagonal matrix while
the symbol $*$ is used to represent symmetric entries in symmetric matrices.

\vspace{-0.2cm}
\section{The Linearized DAE Model of Power Networks}\label{sec:modeling}
We consider a power network consisting $N$ number of buses, modeled by a graph $(\mathcal{N},\mathcal{E})$, where $\mathcal{N}= \{1,\ldots, N\}$ is the set of nodes and $\mathcal{E} \subseteq \mathcal{N}\times\mathcal{N}$ is the set of edges. It is then straightforward to define $\mathcal{N}_i\subseteq \mathcal{N}$ as the set of buses that are connected to bus $i$. The set of buses equipped with PMUs are denoted by $\mathcal{N}_M\subseteq\mathcal{N}$.
Note that $\mathcal{N}$ consists of generator and load buses, i.e.,  $\mathcal{N} = \mathcal{G} \cup \mathcal{L}$ where $\mathcal{G} = \{1,\ldots, G\}$ collects the buses containing $G$ synchronous generators while $\mathcal{L} = \{1,\ldots, L\}$ collects the $L$ buses that contain loads only. Note that for bus $i\in \mathcal{G}$, it might also contain a load.
 {Throughout the paper, complex bus voltages are expressed in rectangular forms---that is, $\m v_i := [v_{Ri}\;\;v_{Ii}]^\top$ for $i\in\mc{N}$.}   

\vspace{-0.2cm}
\subsection{Synchronous Generator Model \& Power Flow Equations}\label{ssec:gen_model}
In this work, we leverage the standard \nth{4}-order dynamics of a two-axis, transient model of synchronous generators, which excludes turbine governor and exciter dynamics. 
The dynamics of a synchronous generator $i \in \mathcal{G}$ can be written as~\cite{sauer2017power}
\begin{subequations} \label{eq:SynGen}
	\begin{align}
	\dot{\delta}_{i} &= \omega_{i} - \omega_{0} \label{eq:SynGen1} \\ 
	\begin{split}
	M_{i}\dot{\omega}_{i} &= T_{{M}i}-\left(e'_{qi}-x'_{di}i_{di}\right)i_{qi}-\left(e'_{di}-x'_{qi}i_{qi}\right)i_{di}\\
	&\quad - D_{i}(\omega_{i}-\omega_{0}) \end{split}\label{eq:SynGen2}    \\ 
	T'_{d0i}\dot{e}'_{qi} &= -e'_{qi} - \left(x_{di} - x'_{di} \right)i_{di} + E_{fdi}  \label{eq:SynGen3} \\
	T'_{q0i}\dot{e}'_{di} &= -e'_{di} + \left(x_{qi} - x'_{qi} \right)i_{qi}. \label{eq:SynGen4}  
	\end{align} 
\end{subequations}
The time-varying parts in \eqref{eq:SynGen} include $\delta_{i}:=\delta_i(t)$ denotes the generator rotor angle, $\omega_{i}:=\omega_i(t)$ denotes the generator rotor speed, $e'_{qi}:=e'_{qi}(t)$ and  $e'_{di}:=e'_{di}(t)$ are the transient voltages along $\mathrm{q}$ and $\mathrm{d}$ axes, $T_{{M}i} := T_{{M}i}(t)$ denotes the generator's mechanical input torque, $E_{fdi} := E_{fdi}(t)$ denotes the internal field voltage, while $i_{qi}$ and $i_{di}$ are generator's stator currents at $\mathrm{q}$ and $\mathrm{d}$ axes. The constants in \eqref{eq:SynGen} are as follows: $M_i$ is the rotor's inertia constant ($\mr{pu} \times \mr{s}^2$), $D_i$ is the damping coefficient ($\mr{pu} \times \mr{s}$), $x_{qi}$ and $x_{di}$ are the direct-axis synchronous reactance ($\mr{pu}$), $x'_{qi}$ and $x'_{di}$ are the direct-axis transient reactance ($\mr{pu}$), $T'_{q0i}$ and $T'_{d0i}$ are the open-circuit time constants ($\mr{s}$), and $\omega_{0}$ denotes the rotor's synchronous speed ($\mathrm{rad/s}$). 
Note that each synchronous generator has a total of four dynamic states, given by $\tilde{\m x}_{{i}} :=[\delta_{i} \;\; \omega_{i} \;\; e'_{qi} \;\;  e'_{di} ]^{\top}$, whereas  generator's mechanical input torque and internal field voltage are considered as inputs, i.e.,  $\tilde{\m u}_{{i}} := [T_{{M}i} \;\; E_{fdi}]^{\top} $. Note that these inputs may not be available for measurement.
The relations among generator's rotor angle, transient voltage, stator's current, and terminal voltage are represented by
the stator electrical circuit, which can be described by the following algebraic constraints~\cite{sauer2017power}
\begin{subequations}\label{eq:SynGenStator}
	\begin{align}
	e'_{di} - v_{Ri}\sin\delta_i+ v_{Ii}\cos\delta_i-R_{si}i_{di}+x'_{qi}i_{qi} &= 0 \label{eq:SynGenStator1} \\
	e'_{qi} - v_{Ri}\cos\delta_i- v_{Ii}\sin\delta_i-R_{si}i_{qi}-x'_{di}i_{di} &= 0,\label{eq:SynGenStator2}
	\end{align}
\end{subequations}
where $R_{si}$ indicates stator's resistance ($\mr{pu}$). 
We define $\m i_{gi}(t)$ as generator's stator currents constructed as $\m i_{gi} := [i_{di} \,\, i_{qi}]^{\top}$. Together, \eqref{eq:SynGen} and \eqref{eq:SynGenStator} form the generator's DAE. 

The power flow equations resembling the power balance between generators and loads can be expressed as~\cite{sauer2017power}  
\begingroup
\allowdisplaybreaks 
\begin{subequations} \label{eq:GPF}
	\begin{align} 
	\begin{split}
&P_{G,i} + 	P_{L,i} = G_{ii}\left(v_{Ri}^2+v_{Ii}^2\right)\\
&+\sum_{j\neq i}^{N}\hspace{-0.05cm}G_{ij}\hspace{-0.05cm}\left(v_{Ri}v_{Rj}\hspace{-0.05cm}+\hspace{-0.05cm}v_{Ii}v_{Ij}\right)\hspace{-0.05cm}+\hspace{-0.05cm}B_{ij}\hspace{-0.05cm}\left(v_{Ii}v_{Rj}\hspace{-0.05cm}-\hspace{-0.05cm}v_{Ri}v_{Ij}\right) 
\end{split}\label{eq:GPF1}\\
	\begin{split}
	&Q_{G,i} + 	Q_{L,i} = -B_{ii}\left(v_{Ri}^2+v_{Ii}^2\right)\\ &-\sum_{j\neq i}^{N}\hspace{-0.05cm}G_{ij}\hspace{-0.05cm}\left(v_{Ri}v_{Ij}\hspace{-0.05cm}-\hspace{-0.05cm}v_{Ii}v_{Rj}\right)\hspace{-0.05cm}+\hspace{-0.05cm}B_{ij}\left(v_{Ri}v_{Rj}\hspace{-0.05cm}+\hspace{-0.05cm}v_{Ii}v_{Ij}\right),\end{split}\label{eq:GPF2}		
	\end{align}
\end{subequations}
\endgroup
for all generator bus $i \in \mathcal{G}$ whereas $G_{ij}$ and $B_{ij}$ respectively denote the conductance and susceptance between bus $i$ and $j$. In the left-hand sides of \eqref{eq:GPF1} and \eqref{eq:GPF2}, $P_{L,i}$ and $Q_{L,i}$  denote the real and reactive load power, which can be time-varying and nonlinearly related with the bus voltage $\m v_i$\cite{sauer2017power}, where $P_{G,i}$ and $Q_{G,i}$ are generator's real and reactive power, computed as
\begin{subequations}\label{eq:GenPower}
	\begin{align}
	\hspace{-0.25cm}P_{G,i} &= \left(v_{Ri}i_{di}+ v_{Ii}i_{qi}\right)\sin\delta_i + \left(v_{Ri}i_{qi}-v_{Ii}i_{di}\right)\cos\delta_i \\
	\hspace{-0.25cm}Q_{G,i} &= \left(v_{Ri}i_{di} + v_{Ii}i_{qi}\right)\cos\delta_i + \left(v_{Ii}i_{di}- v_{Ri}i_{qi}\right) \sin\delta_i.
	\end{align}
\end{subequations}
The power flow equations for load buses $i \in \mathcal{L}$ are the same as in \eqref{eq:GPF} but with an exception that $P_{G,i} = 0$ and $Q_{G,i} = 0$. 
The multi-machine power networks \eqref{eq:SynGen}-\eqref{eq:GenPower} can be lumped into
\begin{align}
\dot{\tilde{\m x}}(t) &= \m f(\tilde{\m x}(t),\m i_g(t),\tilde{\m u}(t)),\;\; 
\m 0 = \m g(\tilde{\m x}(t),\m i_g(t),\m v(t)),\label{eq:nonlinearDAE}
\end{align}
where the nonlinear mapping $\m f:\mbb{R}^{4G}\times\mbb{R}^{2G}\times\mbb{R}^{2G}\rightarrow\mbb{R}^{4G}$ and $\m g:\mbb{R}^{4G}\times\mbb{R}^{2G}\times\mbb{R}^{2N}\rightarrow\mbb{R}^{2G+2N}$
represent the differential and algebraic equations for the power networks. The vectors $\tilde{\m x}\hspace{-0.05cm}:=\hspace{-0.05cm}\{\tilde{\m x}_i\}_{i\in \mc{G}}\hspace{-0.05cm}\in\hspace{-0.05cm}\mbb{R}^{4G}$, ${\m i_g}\hspace{-0.05cm}:=\hspace{-0.05cm}\{\m i_{g,i}\}_{i\in \mc{G}}\hspace{-0.05cm}\in\hspace{-0.05cm}\mbb{R}^{2G}$, $\tilde{\m u}\hspace{-0.05cm}:=\hspace{-0.05cm}\{\tilde{\m u}_i\}_{i\in \mc{G}}\hspace{-0.05cm}\in\hspace{-0.05cm}\mbb{R}^{2G}$, ${\m v}\hspace{-0.05cm}:=\hspace{-0.05cm}\{{\m v}_i\}_{i\in \mc{N}}\hspace{-0.05cm}\in\hspace{-0.05cm}\mbb{R}^{2N}$ represent generators' internal states, stator currents, inputs, and power network's bus voltages. In \eqref{eq:nonlinearDAE}, $\m i_g$ and $\m v$ act as algebraic variables.

\vspace{-0.4cm}
\subsection{Power Network's Linearized DAE Model}\label{ssec:model_linearization}  
 The linearization of power networks \eqref{eq:nonlinearDAE} around a particularly stable operating point $\delta_i^0$, $\omega_i^0$, $e^{\prime 0}_{qi}$ $e^{\prime 0}_{di}$, $i_{di}^0$, $i_{qi}^0$, $T_{{M}i}^0$, $E_{fdi}^0$ for $i \in \mathcal{G}$, $P_{L,i}^0$ and $Q_{L,i}^0$ for $i \in \mathcal{L}$, and $\m v_i^0$ for $i \in \mathcal{N}$ are performed as follows. First, the synchronous generator's linearized dynamics model can be computed and given as 
%\vspace{-0.1cm}
\begin{align}
\m E_D\Delta \dot{\tilde{\m x}}(t) = \m A_{D}\Delta \tilde{\m x} (t) + \m D_{D} \Delta {\m i}_{g}(t) + \m B_{D} \Delta \tilde{\m u}(t).\label{eq:SynGenLin}
\end{align}
In \eqref{eq:SynGenLin}, $\Delta {\tilde{\m x}} := \tilde{\m x}-\tilde{\m x}^0$, $\Delta {\m i}_{g} := {\m i}_{g} - {\m i}_{g}^0$, and $\Delta \tilde{\m u} := \tilde{\m u}-\tilde{\m u}^0$. The linearized stator's algebraic equations, with $\Delta {\m v}_{g} := {\m v}_{g} - {\m v}_{g}^0$ in which ${\m v}_{g} := \{\m v_i\}_{i\in \mc{G}}$ collects the voltages of all generator buses, can be expressed as
%\vspace{-0.1cm}
\begin{align}
0 = \m A_{A} \Delta \tilde{\m x}(t) + \m D_{A} \Delta {\m i}_{g}(t) + \m G_{A} \Delta {\m v}_{g}(t). \label{eq:SynGenStatorLin}
\end{align} 
Finally, the linearized power flow equations are given as
\begin{subequations}\label{eq:PFLin}
%	\vspace{-0.1cm}
\begin{align}
	\begin{split}
0 &= \m A_{G} \Delta \tilde{\m x}(t) + \m D_{G} \Delta {\m i}_{g}(t) + \m G_{GG} \Delta {\m v}_{g}(t)\\ &\quad+ \m G_{GL} \Delta {\m v}_{l}(t) \end{split}\label{eq:PFLin1}\\
0 &= \m G_{LG} \Delta {\m v}_{g}(t)+ \m G_{LL} \Delta {\m v}_{l}(t), \label{eq:PFLin2}
\end{align}
\end{subequations}
where $\Delta {\m v}_{l} := {\m v}_{l} - {\m v}_{l}^0$ and ${\m v}_{l} := \{\m v_i\}_{i\in \mc{L}}$ populates the voltages of all load buses. {Readers are referred to Appendix \ref{appdx:A} for the details on the matrices above.} It is worth noting that the form of the matrices in \eqref{eq:PFLin} depends on the type of the loads---whether they are of constant impedance, current, or power \cite{sauer2017power}. Now, from \eqref{eq:SynGenStatorLin}, we have $\Delta {\m i}_{g}(t)= -\m D_{A}^{-1}\m A_{A} \Delta \tilde{\m x}(t)-\m D_{A}^{-1}\m G_{A} \Delta {\m v}_{g}(t)$. % assuming that the matrix $\m D_{A}$ is nonsingular. 
Note that the premise on the nonsingularity of $\m D_{A}$ is not restrictive since, by the construction of $\m D_{A}$ provided in Appendix \ref{appdx:A}, it follows that $\m D_{A}^{-1}$ exists if and only if
\begin{align*}
R_{si}^2 + x'_{di}x'_{qi} &\neq 0,\quad \forall i\in\mc{G}.
\end{align*}
Substituting the expression for $\Delta {\m i}_{g}$ into \eqref{eq:SynGenLin} and \eqref{eq:PFLin} yields
\begin{align}\label{eq:linearDAEsimple}
\underbrace{\bmat{\m E_{D}\;\m O \\ \m O \;\;\; \m O}}_{\m E}\hspace{-0.05cm}\underbrace{\bmat{\Delta \dot{\tilde{\m x}}(t) \\ \Delta \dot{\m v}(t)}}_{\dot{\m x}(t)} \hspace{-0.1cm}=\hspace{-0.1cm}\underbrace{\bmat{\m A_1 \;\; \m A_2 \\ \m A_3 \;\; \m A_4}}_{\m A}\hspace{-0.05cm}\underbrace{\bmat{\Delta \tilde{\m x} (t) \\ \Delta {\m v} (t)}}_{{\m x}(t)}\hspace{-0.05cm}+\hspace{-0.05cm}\underbrace{\bmat{\m{B}_D\\ \m O}}_{\m B_u}\hspace{-0.05cm}{\m u}(t),
\end{align}
where in \eqref{eq:linearDAEsimple}, $\Delta \m v^\top := \bmat{\Delta {\m v}_{g}^\top \;\;\Delta {\m v}_{l}^\top}$, $\m u := \Delta \tilde{\m u}$, and
\begin{align*}
\m A_1 &= \m A_{D}-\m D_{D}\m D_{A}^{-1}\m A_{A},\;
\m A_2 = \bmat{ -\m D_{D}\m D_{A}^{-1}\m G_{A}&\m O},\\
\m A_3 &= \bmat{\m A_{G}-\m D_{G}\m D_{A}^{-1}\m A_{A} \\ \m O},\;
\m A_4 = \bmat{\bar{\m G} & \m G_{GL} \\ \m G_{LG} & \m G_{LL}},
\end{align*}
with $\bar{\m G}$ in $\m A_4$ is equal to $\m G_{GG}-\m D_{G}\m D_{A}^{-1}\m G_{A}$. From \eqref{eq:linearDAEsimple}, a compact linear DAE for power networks can be written in a descriptor form $\m E\dot{\m x}(t) = \m A {\m x}(t) + \m B_u \m u(t)$, where $\m E\in\mbb{R}^n$, $\m A\in\mbb{R}^n$, $\m B_u\in\mbb{R}^m$ are constant matrices with $n = 4G+2N$ and $m = 2G$, $\m E$ is singular with $\rank (\m E) := r$ where $r$ equals to $\rank(\m E_D) = 4G$, ${\m x}\in\mbb{R}^n$ comprises differential and algebraic variables $\Delta\tilde{\m x}$  and $\Delta {\m v}$ such that ${\m x}^\top = \bmat{\Delta\tilde{\m x}^\top\;\;\Delta {\m v}^\top}$, and $\m u\in\mbb{R}^m$ denotes the deviation of control inputs around their steady-state values. The measurement model for linear DAE \eqref{eq:linearDAEsimple} with PMUs is provided in the next section.

\vspace{-0.4cm}
\subsection{Power Network's Measurement Model with PMUs}\label{ssec:PMU_model}
For a PMU installed on bus $j\in\mathcal{N}_M$, it is practical to consider that it has the capability to measure \textit{(i)} bus voltage $\m v_j$ and \textit{(ii)} line currents for all branches that are connected to bus $j$ \cite{gomez2011use,Risbud2016,Kekatos2012PMU,Zhou2006PMU}. Define $\mathcal{N}_j\subseteq \mathcal{N}$ as the set of buses connected to bus $j$. {A line current $\m I_{jk} = I_{Rjk}+ jI_{Ijk}$ constitutes the current flowing on a branch originating from bus $j$ to bus $k$. The set of line currents measured by a PMU at bus $j\in\mathcal{N}_M$  is defined as $\mathcal{I}_j:=\{\m I_{jk}\}_{k\in\mathcal{N}_{j}}$ such that $|\mathcal{I}_j| = |\mathcal{N}_{j}|$. To obtain the expressions for line currents, we need to utilize the branch admittance matrix $\m Y_{ft}$ which is constructed as 
\begin{align}
\m Y_{ft} := \bmat{\m Y_{f} \\ \m Y_{t}}\in\mathbb{C}^{2|\mathcal{E}|\times|\mathcal{N}|}, \label{eq:Ybranch}
\end{align}
where $\m Y_{f}\in\mathbb{C}^{|\mathcal{E}|\times|\mathcal{N}|}$ and $\m Y_{t}\in\mathbb{C}^{|\mathcal{E}|\times|\mathcal{N}|}$ are the \textit{from} and \textit{to} branch admittance matrices \cite{MATPOWER2011}.} As such, all line currents can be 
computed using the method provided in \cite{Risbud2016}. {That is
\begin{align}
\bmat{\mathcal{I}_R\\ \mathcal{I}_I} = \bmat{\mathrm{Re}(\m Y_{ft})& -\mathrm{Im}(\m Y_{ft}) \\ \mathrm{Im}(\m Y_{ft})& \mathrm{Re}(\m Y_{ft})}\tilde{\m v},\label{eq:linecurrents}
\end{align}
where $\tilde{\m v}^\top := \bmat{\{v_{Ri}\}_{i\in \mc{N}}^\top\;\;\{v_{Ii}\}_{i\in \mc{N}}^\top}$ is the rearranged network's bus voltages (since $\tilde{\m v}$ has different ordering from ${\m v}$ given in Section \ref{ssec:gen_model}).}
This configuration allows a PMU installed on bus $j\in \mathcal{N}_M$ to  measure the following quantities
\begin{align*}
\m y_j^\top = \bmat{v_{Rj}&v_{Ij}&\{I_{Rjk}\}_{k\in\mathcal{N}_{j}}^\top&\{I_{Ijk}\}_{k\in\mathcal{N}_{j}}^\top}, %\label{eq:PMU2}
\end{align*}
where each $I_{Rjk}$ and $I_{Ijk}$ can be linearly obtained from bus voltages via the utilization of $\m Y_{ft}$, thanks to \eqref{eq:linecurrents}, provided as $\m y_j(t) = \tilde{\m C}_j\tilde{\m v}(t)$ where $\tilde{\m C}_j$ is given as \cite{Risbud2016}
	\begin{align*}
	\tilde{\m C}_j := \bmat{\m e_j^\top & \m 0 \\ \m 0 & \m e_j^\top\\ \m S_j\mathrm{Re}(\m Y_{ft})& -\m S_j\mathrm{Im}(\m Y_{ft})\\\m S_j\mathrm{Im}(\m Y_{ft})& \m S_j\mathrm{Re}(\m Y_{ft})}.%\label{eq:PMU_measurement2} 
	\end{align*}
In the above, $\m e_j$ is the vector of standard basis in $\mathbb{R}^N$ with $1$ at row $j$ and zero otherwise and $\m S_j\in\mathbb{R}^{|\mathcal{N}_j|\times2|\mathcal{E}|}$ is a binary, selection matrix that selects the corresponding row of $\m Y_{ft}$ which are originating from bus $j$. The overall measurement model for power networks with PMUs can then be expressed as
\begin{align}
\m y(t) = \tilde{\m C}\m C_M{\m v}(t), \label{eq:PMU_measurement_all} 
\end{align}
where in \eqref{eq:PMU_measurement_all}, the vector $\m y\in\mbb{R}^p$ lumps the measured outputs, $\tilde{\m C} = \{\tilde{\m C}_{i}\}_{i\in \mathcal{N}_M}$, and $\m C_M$ is an orthogonal matrix such that $\tilde{\m v}(t) = \m C_M{\m v}(t)$. The branch admittance matrix $\m Y_{ft}$ in \eqref{eq:Ybranch} can be constructed from MATPOWER \cite{MATPOWER2011}. 

\vspace{-0.3cm}
\section{Regularity, Index Characterization, and Observability of Power Networks }\label{sec:reg_idx_obs}
The linearized power network's DAE \eqref{eq:linearDAEsimple} with measurement \eqref{eq:PMU_measurement_all} can be lumped into the following state-space equations  
	\begin{align}
	\Aboxed{\m E\dot{\m x}(t) &= \m A {\m x}(t) + \m B_u \m u(t), \;\quad\m y(t) = \m C{\m x}(t),} \label{eq:DAE_base} 
	\end{align}
where $\m C = \bmat{\m O \;\;\; \tilde{\m C}\m C_M}$. 
As opposed to ODEs, the existence and uniqueness of solutions for DAEs are not always guaranteed for any initial conditions. 
On that regard, one of the most important property for DAEs is \textit{regularity}. That is, a linear DAE is said to be regular if and only if it has a unique solution for every consistent initial condition \cite{duan2010analysis}. The regularity of DAE can be characterized from the matrix pair $(\m E,\m A)$, i.e., the DAE \eqref{eq:DAE_base} is regular if and only if there exists $s\in\mbb{C}$ such that $\det{(s\m E-\m A)}\neq 0$. For the case of power networks \eqref{eq:DAE_base}, first define $\tilde{\Lambda}(\cdot)$ as the set of finite poles of a linear DAE pair
\begin{align*}
\tilde{\Lambda}(\m E,\m A) := \{s\in\mbb{C}\,|\,\det(s\m E-\m A) = 0\}. 
\end{align*}
Now, the regularity condition of power networks becomes
	\begin{align}
\det{(s\m E-\m A)} &= \det\left(\bmat{s\m E_D - \m A_1 & -\m A_2 \\ -\m A_3 & -\m A_4}\right).\label{eq:DAE_base_reg_1} 
	\end{align}
By considering a finite $s\notin \tilde{\Lambda}(\m E,\m A)$ such that $s\m E_D - \m A_1$ is nonsingular, the right-hand side of \eqref{eq:DAE_base_reg_1} is equivalent to
	\begin{align}
\det\left(\m A_4 + \m A_3\left(s\m E_D - \m A_1\right)^{-1} \m A_2 \right)\neq 0.\label{eq:DAE_base_reg_2} 
\end{align} 	 
Hence, the DAE \eqref{eq:DAE_base} is regular if and only if \eqref{eq:DAE_base_reg_2} holds for some finite $s\notin \tilde{\Lambda}(\m E,\m A)$. It is difficult to assess the regularity of DAE \eqref{eq:DAE_base} due to weak structure in \eqref{eq:DAE_base} or lack thereof. However, several researches have been conducted to analyze the solvability of simplified power network's DAEs---see \cite{Gross2014,Gross2016}. It is discovered therein that, for a lossless power network with unity magnitude on all bus voltages, the regularity of the linearized DAE model is ensured if and only if there exist paths such that every load is connected to a generator bus.  

Another important property in linear DAE is the \textit{differentiation index}, which refers to the number of differentiations that need to be executed in order to transform the DAE into an ODE. 
 It is understood that the DAE model of the aforementioned simplified power networks is of index one if only if it is regular \cite{Gross2014,Gross2016}. {For power networks model considered in this paper, it is revealed that if $\m A_4$ is of full rank, then DAE \eqref{eq:DAE_base} is impulse-free\footnote{Linear DAEs that are {impulse-free} do not contain impulsive terms in their solutions for given arbitrary initial conditions.}, of index one, and regular---shown in the result below (the proof is provided in  Appendix \ref{appdx:B}).} 
\vspace{-0.15cm}
\begin{theorem}\label{thm:regular}
	Suppose that the matrix $\m A_4$ in DAE \eqref{eq:DAE_base} is of full rank. Then, the power networks DAE \eqref{eq:DAE_base} is impulse-free, of index one, and regular.
\end{theorem}
\vspace{-0.15cm}
\noindent The above proposition provides another alternative other than \eqref{eq:DAE_base_reg_2} to determine the regularity of the power networks model. Because of its useful implications, the following statement is assumed throughout the paper.
\vspace{-0.15cm}
\begin{asmp}\label{asmp:regular}
The matrix $\m A_4$ in DAE \eqref{eq:DAE_base} is of full rank.
\end{asmp}
\vspace{-0.15cm}
\noindent 
{The matrix $\m A_4$ describes the relation among loads and generators via power flow equations. Thus, the nonsingularity of $\m A_4$ simply translates to the connectivity among generators and loads, which is similar to the results presented in \cite{Gross2014,Gross2016} on the regularity of the DAE.
Note that Assumption \ref{asmp:regular} is mild and holds for the IEEE test cases considered in the numerical study section. } 

Before a state observer can be designed, it is crucial to assess whether the DAE's internal states can be estimated via limited measurements. There exist numerous concepts for observability in DAEs, in addition to detectability---see \cite{duan2010analysis}. 
{However, as we are looking to design observers with the \textit{least} possible assumptions, we choose herein to consider detectability and impulse observability---discussed as follows.}
The DAE \eqref{eq:DAE_base} is said to be \textit{detectable} if and only if there exists a matrix $\m L\in \mbb{R}^{n\times p}$ such that the DAE described by the matrix pair $(\m E,\m A - \m L \m C)$ is stable, that is, $\tilde{\Lambda}(\m E,\m A - \m L \m C)\in {\mbb{C}}_-$ or equivalently, all finite eigenvalues of $(\m E,\m A - \m L \m C)$ are on open left-half of complex plane. The detectability of DAE \eqref{eq:DAE_base} can be ensured if the following rank condition holds \cite{duan2010analysis}
\begin{align}
\rank \left(\bmat{s \m E -\m A \\ \m C}\right) = n,\;\;\;\;\forall s \in \bar{\mbb{C}}_+,\label{eq:detectability}
\end{align}
which is akin to the PBH test for detectability of linear differential systems. 
On the other hand, the impulse observability (or \textit{I-observability})  condition concerns with the ability to observe impulsiveness of the actual states from the impulsive behavior of the measurements. 
The condition that ensures {I-observability} of a linear DAE is given by \cite{duan2010analysis}   
\begin{align}
\rank \left(\bmat{\m E & \m A \\ \m O & \m E \\ \m O & \m C}\right) &= n + \rank(\m E). \label{eq:I-observability}
\end{align}
The following result presents a reduced equivalent condition to achieve I-observability for power network's DAE.
\vspace{-0.15cm} 
\begin{theorem}\label{thm:I-observability}
The DAE \eqref{eq:DAE_base} is I-observable if and only if 
\begin{align}
\hspace{-0.2cm}\rank\hspace{-0.05cm}\left(\hspace{-0.05cm}\bmat{\;\m A_3 \;\;\;\;\, \m A_4 \;\;\;\\ \m E_D \;\;\;\;\;\m O\;\;\;\; \\ \;\m O \;\;\;\;\tilde{\m C}\m C_M}\hspace{-0.05cm}\right)\hspace{-0.1cm} =\hspace{-0.05cm}\rank\hspace{-0.05cm}\left(\hspace{-0.05cm}\bmat{\m A_4^{-1}\m A_3 \;\; \m C_M^\top \\ \;\;\;\,\m I \;\;\;\;\;\;\;\,\m O \;\;\;\\\ \;\;\m O \;\;\;\;\;\;\;\tilde{\m C}\;\;\;}\hspace{-0.05cm}\right) \hspace{-0.1cm}&= \hspace{-0.00cm}n. \label{eq:I-observability-dae}
\end{align}	
\end{theorem}
\vspace{-0.10cm}
\noindent Please refer to Appendix \ref{appdx:C} for the proof of Theorem \ref{thm:I-observability}. On that regard, the following assumption is considered.
\vspace{-0.15cm}
\begin{asmp}\label{asmp:detect_impulse_obs}
	The PMUs are distributed in such a way that the power network's DAE \eqref{eq:DAE_base} is detectable and I-observable, i.e., \eqref{eq:detectability} and \eqref{eq:I-observability-dae} are satisfied.
\end{asmp}
\vspace{-0.15cm}
\noindent 
{The above assumption is crucial for enabling power system's DSE using observers---detailed in Section \ref{sec:DAE_observer}. 
It is worth noting that this assumption is also mild and easily satisfied since we use similar number and location of PMUs as in the literature.	
	Regardless, the problem of placing PMUs in order to achieve detectability and I-observability is not covered in this work as this problem deserves a standalone dedication and an in-depth analysis which are beyond the scope of this paper and as such, is left for future research.}
In the following section, we explore several DAE state observer designs that can be utilized to perform DSE based on DAE \eqref{eq:DAE_base}. 

\vspace{-0.2cm}
\section{Synthesis of Luenberger DAE State Observers}\label{sec:DAE_observer}
 There exist two different realizations of state observer for DAE: DAE observer and ODE observer. The former is considered herein mainly due to \textit{(i)} its simplicity as no equivalence transformation required and \textit{(ii)} it only assumes detectability and I-observability. This is in contrast to many ODE observers where the original DAE needs to be transformed into a specific restricted system equivalent while requiring more stricter conditions, e.g. in \cite{Gupta2014}.  
With that in mind, two kinds of DAE observer are presented. 
in this section: \textit{(a)} a simple Luenberger observer and \textit{(b)} a robust $\mc{H}_{\infty}$ observer. 
First, based on \cite{Luenberger1971}, the next Luenberger DAE state observer for \eqref{eq:DAE_base} is proposed 
\vspace{-0.0cm}
\begin{subequations}\label{eq:DAE_Luenberger_obs}
	\begin{empheq}[box=\widefbox]{align}
\m E\dot{\hat{\m x}}(t) &= \m A \hat{\m x}(t) + \m B_u \m u(t) + \m L(\m y(t) - \hat{\m y}(t)) \label{eq:DAE_Luenberger_obs_1}\\ 
\hat{\m y}(t) &= \m C\hat{\m x}(t),\label{eq:DAE_Luenberger_obs_2}
	\end{empheq}
\end{subequations}
where $\hat{\m x}\in\mbb{R}^n$ denotes the estimate of the actual system's state ${\m x}$, $\hat{\m y}\in\mbb{R}^p$ is the output estimate, and $\m L\in \mbb{R}^{n\times p}$ is the observer gain matrix. Our goal is finding $\m L$ such that $
\lim_{t\rightarrow\infty}(\m x(t)-\hat{\m x}(t)) = \m 0$ for arbitrary initial states $\m x_0$, $\hat{\m x}_0$  and inputs $\m u(t)$ for all $t \geq 0$. If the state estimation error is defined as $\m e(t) := \m x(t)-\hat{\m x}(t)$, the error dynamics can be written as 
\begin{align}
\m E\dot{{\m e}}(t) &= (\m A -\m L \m C)\m e(t). \label{eq:DAE_error_dynamics}
\end{align} 
Note that the existence of $\m L$ such that \eqref{eq:DAE_error_dynamics} is stable is guaranteed due to the detectability condition in Assumption \ref{asmp:detect_impulse_obs}. 
A linear DAE that is impulse-free, regular, and stable is called \textit{admissible} \cite{xu2006robust,Dai1989}. To that end,  we seek for a systematic way to compute $\m L$ that makes the error dynamics in \eqref{eq:DAE_error_dynamics} to be admissible. The following result from \cite{xu2006robust} provides a sufficient and necessary condition for the  admissibility of DAE \eqref{eq:DAE_error_dynamics}. 
\vspace{-0.15cm}
\begin{mylem}[\hspace{-0.012cm}\cite{xu2006robust}]\label{lem:admissibility}
A linear DAE denoted by $(\tilde{\m E},\tilde{\m A})$ is admissible if and only if there exist $\m X \in\mbb{S}^n_{++}$ and $\m Y\in\mbb{R}^{(n-r)\times n}$ such that the following LMI is feasible
\begin{align}
\tilde{\m A}^{\top}(\m X \tilde{\m E} + \tilde{\m E}^{\perp\top} \m Y) + (\m X \tilde{\m E} + \tilde{\m E}^{\perp\top} \m Y)^\top\tilde{\m A} \prec 0,\label{eq:admissibility}
\end{align}
where $\tilde{\m E}^{\perp}\in\mbb{R}^{(n-r)\times n}$ is an orthogonal complement of $\tilde{\m E}$.
\end{mylem}
\vspace{-0.15cm}
\noindent {Using \eqref{eq:admissibility}, a condition for admissibility of state estimation error dynamics is obtained---see  Appendix \ref{appdx:D} for the proof.}
\vspace{-0.15cm}
\begin{theorem}\label{thm:admissibility_error_dyn}
The state estimation error dynamics in \eqref{eq:DAE_error_dynamics} is admissible if and only if there exist $\m X \in\mbb{S}^n_{++}$, $\m Y\in\mbb{R}^{(n-r)\times n}$, and $\m W\in\mbb{R}^{n\times p}$ such that the following LMI is feasible
\begin{align}
\m A ^{\top} \m X \m E + \m E^\top\m X\m A + \m A ^{\top} &\m E^{\perp\top} \m Y + \m Y^\top\m E^{\perp}\m A \nonumber \\
&- \m C^{\top}\m W -\m W^\top \m C \prec 0, \label{eq:admissibility_LMI}
\end{align}
where the gain matrix $\m L$ can be recovered as $\m L = \left(\m W \m P^{-1}\right)^{\top}$ and ${\m E}^{\perp}\in\mbb{R}^{(n-r)\times n}$ is an orthogonal complement of ${\m E}$.
\end{theorem}
\vspace{-0.2cm}
\begin{myrem}\label{rem:optimization_problems}
Although the satisfaction of LMI \eqref{eq:admissibility_LMI} ensures the stability of error dynamics \eqref{eq:DAE_error_dynamics}, the convergence rate to which  $\hat{\m x}(t)$ approaches ${\m x}(t)$ as $t \rightarrow \infty$ can be relatively poor. As a potential remedy, one can minimize the maximum eigenvalue of $\m E^\top \m X \m E$ by solving the following convex optimization problem
\begin{align*}
	{\mathbf{P1}}\min_{\kappa, \m X, \m Y, \m W} \kappa \;;\;
	\sto \, \eqref{eq:admissibility_LMI},\, \kappa > 0, \m X \succ 0,\kappa \m I - \m E^\top \m X \m E \succeq 0. 
\end{align*}
\end{myrem} 
\vspace{-0.1cm}
\noindent Theorem \ref{thm:admissibility_error_dyn} provides a systematic method to find  $\m L$ that makes  \eqref{eq:DAE_error_dynamics} admissible as the LMI in \eqref{eq:admissibility_LMI} can be solved using standard semidefinite programming (SDP) solvers.

The above Luenberger DAE state observer \eqref{eq:DAE_Luenberger_obs} assumes an ideal operating condition of power networks in the sense that no disturbance is present. Nevertheless, in a more practical set up, disturbances can manifest in various forms: from unknown inputs, process and measurement noise, to cyber-attacks. Hence, in addition to the the Luenberger observer \eqref{eq:DAE_Luenberger_obs}, here we also design a robust $\mc{H}_{\infty}$ DAE observer to minimize the effect of disturbance to the state estimation error. The power network's DAE subject to external disturbances can be written as 
\begin{subequations}\label{eq:DAE_base_dist} 
	\begin{empheq}[box=\widefbox]{align}
	\m E\dot{\m x}(t) &= \m A {\m x}(t) + \m B_u \m u(t) + \m B_w \m w(t) \label{eq:DAE_base_dist_1} \\
	\quad\m y(t) &= \m C{\m x}(t) + \m D_w \m w(t), \label{eq:DAE_base_dist_2} 
\end{empheq}
\end{subequations}
where $\m w\in\mathbb{R}^q$ lumps all disturbances into a single vector and matrices $\m B_w$ and $\m D_w$ describe how the external disturbances are distributed in the system. Specifically, $\m B_w$ takes into account any disturbances affecting the system's dynamics whereas $\m D_w$ considers disturbances affecting the measurements. It is presumed from now on that matrices $\m B_w$ and $\m D_w$ are known. By utilizing the same Luenberger state observer as in \eqref{eq:DAE_Luenberger_obs}, the estimation error dynamics can be derived as follows
\begin{subequations}\label{eq:DAE_error_dynamics_dist} 
	\begin{align}
	\m E\dot{{\m e}}(t) &= (\m A -\m L \m C)\m e(t) + (\m B_w - \m L \m D_w)\m w(t) \label{eq:DAE_error_dynamics_dist_1} \\
	\quad\m {\epsilon}(t) &= \m \Gamma\m e(t), \label{eq:DAE_error_dynamics_dist_2} 
	\end{align}
\end{subequations}
where $\m {\epsilon}\in\mbb{R}^n$ is the performance of error dynamics with respect to the user-defined performance matrix $\m\Gamma\in\mathbb{R}^{n\times n}$. The next result summarizes the design of a robust $\mc{H}_{\infty}$ DAE observer.
\vspace{-0.2cm}
\begin{theorem}\label{thm:robust_observer} 
For the state estimation error dynamics in \eqref{eq:DAE_error_dynamics_dist}, it holds that \textit{(a)} it is stable whenever $\m w(t) = \m 0$ for all $t \geq 0$ and {(b)}  $\norm{\m \epsilon(t)}^2_{L2} <\gamma \norm{\m w(t)}^2_{L2}$ where $\gamma \geq 0$ for every bounded disturbance $\m w(t)$ and zero initial error $\m e_0 = \m 0$, if there exist $\m X \in\mbb{S}^n_{++}$, $\m Y\in\mbb{R}^{(n-r)\times n}$, and $\m W\in\mbb{R}^{n\times p}$ such that 
\begin{align}
\bmat{\m A ^{\top} \m X \m E + \m E^\top\m X\m A + \m \Xi + \m \Gamma^\top \m \Gamma & * \\ \m B_w^\top \m X\m E + \m B_w^\top\m E^{\perp\top} \m Y - \m D_w^\top \m W & -\gamma \m I} \prec 0, \label{eq:H_infinity}
\end{align}
is feasible where the matrix $\m \Xi$ is defined as
\begin{align*}
\m \Xi := \m A ^{\top} &\m E^{\perp\top} \m Y + \m Y^\top\m E^{\perp}\m A - \m C^{\top}\m W -\m W^\top \m C.
\end{align*}
Moreover, the matrix $\m L$ can be computed as $\m L = \left(\m W \m P^{-1}\right)^{\top}$ and ${\m E}^{\perp}\in\mbb{R}^{(n-r)\times n}$ is an orthogonal complement of ${\m E}$.
\end{theorem}
\vspace{-0.2cm}
{The proof of Theorem \ref{thm:robust_observer} can be obtained from  Appendix \ref{appdx:E}.}
The LMI \eqref{eq:H_infinity} guarantees the boundedness of estimation error performance in the sense that $\norm{\m \epsilon(t)}^2_{L2} <\gamma \norm{\m w(t)}^2_{L2}$ for every bounded disturbance $\m w(t)$ and and zero initial error. To minimize the impact of $\m w(t)$, the following convex problem can be considered
\begin{align*}
	\mathbf{P2}\;\min_{\gamma, \m X, \m Y, \m W}\; \gamma \;;\;
	\sto \; \eqref{eq:H_infinity},\, \gamma > 0,\,\m X \succ 0,
\end{align*}
while the multi-objective optimization problem below can be considered to improve the estimation error convergence rate
\begin{align*}
	\mathbf{P3}\;\min_{\gamma, \kappa,\m X, \m Y, \m W}\;\; &c_1\kappa + c_2 \gamma \;   \\ \subjectto \; &\eqref{eq:H_infinity},\, \gamma,\kappa > 0,\,\m X \succ 0, \,\kappa \m I - \m E^\top \m X \m E \succeq 0,
\end{align*}
where $c_1,\,c_2\in\mbb{R}_{++}$ are predefined constants.

\vspace{-0.2cm}
\section{Tackling Unknown Inputs and Sensor Failures}\label{sec:DAE_observer_UI} 
In generators particularly equipped with brushless excitation systems, it is difficult to measure the exciter's field current and voltage \cite{Ghahremani2011}. Consequently, the field voltage $E_{fd}$ is not always available for the observer \cite{Lee2020,Anagnostou2018,Zhou2013}.  
To that end, this section focuses on extending the DAE state observers introduced in Section \ref{sec:DAE_observer} to consider unknown inputs through a \textit{proportional-integral} (PI) framework. To start, consider the disturbed power networks DAE \eqref{eq:DAE_base_dist} with unknown inputs
\begin{subequations}\label{eq:DAE_base_dist_UI} 
	\begin{align}
	\m E\dot{\m x}(t) &= \m A {\m x}(t) + \m B_u \m u(t) + \m B_{\nu} \m \nu(t) +  \m B_w \m w(t) \label{eq:DAE_base_dist_UI_1} \\
	\quad\m y(t) &= \m C{\m x}(t) + \m D_w \m w(t), \label{eq:DAE_base_dist_UI_2} 
	\end{align}
\end{subequations} 
where $\m \nu\in\mbb{R}^v$ represents unknown inputs and $\m B_{\nu}\in\mbb{R}^{n\times v}$ is a known matrix dictating how the unknown inputs are distributed in the network. 
The idea here is to eliminate the effect of unknown inputs (as well as input sensor failures) by \textit{estimating} the dynamic behavior of the unknown inputs and exploiting the estimate to compensate the state estimation error caused by unknown inputs. An estimate of unknown inputs' dynamics is provided as $\dot{\m \nu}(t) = \m \Psi \m \nu(t)$ for some matrix $\m \Psi$ of appropriate dimension \cite{BAKHSHANDE2015}. The matrix $\m \Psi$ is typically designed based on the knowledge of unknown inputs' dynamics. However, when the dynamics are unknown, the choice $\m \Psi = \m O$ can provide a sufficient estimate \cite{Soffker1995}. The augmented linearized DAEs, which is a reformulation of \eqref{eq:DAE_base_dist_UI}, can be expressed as 
\begin{align*}
\underbrace{\bmat{\m E\;\;\m O \\ \m O \;\;\m I\;}}_{\m E_{\xi}}\hspace{-0.05cm}\underbrace{\bmat{\dot{{\m x}}(t) \\ \dot{\m \nu}(t)}}_{\dot{\m \xi}(t)} \hspace{-0.1cm}&=\hspace{-0.1cm}\underbrace{\bmat{\m A \;\; \m B_{\nu} \\ \m O \;\;\, \m O\;\;}}_{\m A_{\xi}}\hspace{-0.05cm}\underbrace{\bmat{{\m x} (t) \\ {\m \nu} (t)}}_{{\m\xi}(t)}\hspace{-0.05cm}+\hspace{-0.05cm}\underbrace{\bmat{\m{B}_D\\ \m O}}_{\m B_{u,\xi}}\hspace{-0.05cm}{\m u}(t) +\hspace{-0.05cm}\underbrace{\bmat{\m{B}_w\\ \m O}}_{\m B_{w,\xi}}\hspace{-0.05cm}{\m w}(t), \\ %\label{eq:DAE_base_aug_1}\\
\m y(t) &=  \underbrace{\bmat{\m C \;\;\, \m O}}_{\m C_{\xi}}\underbrace{\bmat{{\m x} (t) \\ {\m \nu} (t)}}_{{\m\xi}(t)} +  \m D_w \m w(t), 
\end{align*}
with $\rank\hspace{-0.05cm}\left(\m E_{\xi}\right) \hspace{-0.025cm}=\hspace{-0.025cm} r+\nu$. The above DAE can be reduced into
\begin{subequations}\label{eq:DAE_base_aug} 
	\begin{empheq}[box=\widefbox]{align}
	\m E_{\xi}\dot{\m \xi}(t) &= \m A_{\xi} \m \xi(t) + \m B_{u,\xi} \m u(t) +  \m B_{w,\xi} \m w(t) \label{eq:DAE_base_augI_1} \\
	\quad\m y(t) &= \m C_{\xi}\m \xi(t) + \m D_w \m w(t), \label{eq:DAE_base_aug_2} 
\end{empheq}
\end{subequations} 
where $\m \xi := \bmat{{\m x}^\top \; \m \nu^\top}^\top\hspace{-0.1cm}\in\hspace{-0.05cm}\mbb{R}^{\varsigma}$ is the state vector of the augmented DAE.
The linear DAE \eqref{eq:DAE_base_aug} retains the properties of the power networks DAE \eqref{eq:DAE_base}, which is shown in the result below.
\vspace{-0.2cm}
\begin{theorem}\label{thm:regular_aug}
The augmented linear DAE \eqref{eq:DAE_base_aug} is impulse-free, of index one, and regular. Moreover, the augmented linear DAE \eqref{eq:DAE_base_aug} is detectable and I-observable if and only if the power networks DAE in \eqref{eq:DAE_base} is detectable and I-observable.
\end{theorem}
\vspace{-0.2cm}
{The implication of Theorem \ref{thm:regular_aug} ( Appendix \ref{appdx:F} has the proof) is the existence of an impulse-free DAE state observer for the augmented system \eqref{eq:DAE_base_aug}.} If $\hat{\m \xi} \in \mbb{R}^{\varsigma}$ is an estimate of $\m \xi$ where $\hat{\m \xi}^\top := \bmat{\hat{\m x}^\top \; \hat{\m \nu}^\top}$ and $\hat{\m x},\,\hat{\m \nu}$ are estimates of ${\m x},\,{\m \nu}$ respectively, the PI Luenberger DAE state observer can be constructed as
\begin{align*}
\underbrace{\bmat{\m E\;\;\m O \\ \m O \;\;\m I\;}}_{\m E_{\xi}}\hspace{-0.05cm}\underbrace{\bmat{\dot{\hat{\m x}}(t) \\ \dot{\hat{\m \nu}}(t)}}_{\dot{\hat{\m \xi}}(t)} \hspace{-0.125cm}&=\hspace{-0.125cm}\underbrace{\bmat{\m A \;\; \m B_{\nu} \\ \m O \;\;\, \m O\;\;}}_{\m A_{\xi}}\hspace{-0.05cm}\underbrace{\bmat{\hat{\m x} (t) \\ \hat{\m \nu} (t)}}_{\hat{\m\xi}(t)}\hspace{-0.09cm}+\hspace{-0.09cm}\underbrace{\bmat{\m{B}_D\\ \m O}}_{\m B_{u,\xi}}\hspace{-0.075cm}{\m u}(t)\hspace{-0.07cm} +\hspace{-0.1cm}\underbrace{\bmat{\m L_{\xi,P}\\ \m L_{\xi,I}}}_{\m L_{\xi}}\hspace{-0.075cm}\Delta \m y(t), \\ %\label{eq:DAE_base_aug_1}\\
\hat{\m y}(t) &=  \underbrace{\bmat{\m C \;\;\, \m O}}_{\m C_{\xi}}\underbrace{\bmat{\hat{\m x} (t) \\ \hat{\m \nu} (t)}}_{\hat{\m\xi}(t)}, 
\end{align*}
where $\Delta \m y(t) := \m y(t) - \hat{\m y}(t)$ and $\m L_{\xi}\in \mbb{R}^{\varsigma\times p}$ is the observer gain matrix and comprised of  proportional and integral gains. 
The above observer dynamics can be simplified into 
\begin{subequations}\label{eq:DAE_Luenberger_obs_aug} 
\begin{empheq}[box=\widefbox]{align}
	\m E_{\xi}\dot{\hat{\m \xi}}(t) &= \m A_{\xi} \hat{\m \xi}(t) + \m B_{u,\xi} \m u(t) +  \m L_{\xi} \left(\m y(t) - \hat{\m y}(t)\right) \label{eq:DAE_Luenberger_obs_aug_1} \\
	\quad\hat{\m y}(t) &= \m C_{\xi}\hat{\m \xi}(t). \label{eq:DAE_Luenberger_obs_aug_2} 
	\end{empheq}
\end{subequations} 
Setting $\m e_{\xi} := \m \xi - \hat{\m \xi}$, the state estimation error dynamics can be derived from \eqref{eq:DAE_base_aug} and \eqref{eq:DAE_Luenberger_obs_aug}  and given as
\begin{subequations}\label{eq:DAE_error_dynamics_dist_aug} 
	\begin{align}
		\hspace{-0.2cm}\m E_{\xi}\dot{{\m e}}_{\xi}(t) \hspace{-0.05cm}&= \hspace{-0.05cm}(\m A_{\xi} -\m L_{\xi} \m C_{\xi})\m e_{\xi}(t) + (\m B_{w,\xi} - \m L_{\xi} \m D_w)\m w(t) \label{eq:DAE_error_dynamics_dist_aug_1} \\
		\m {\epsilon}_{\xi}(t)\hspace{-0.05cm} &= \hspace{-0.05cm}\m \Gamma_{\xi}\m e_{\xi}(t). \label{eq:DAE_error_dynamics_dist_aug_2} 
	\end{align}
\end{subequations}
where $\m {\epsilon}_{\xi}\in\mbb{R}^\varsigma$ is the performance of augmented error dynamics and $\m\Gamma_{\xi}\in\mathbb{R}^{\varsigma\times \varsigma}$ is the user-defined performance matrix.
Since the state estimation error dynamics \eqref{eq:DAE_error_dynamics_dist_aug} share a similar structure to that of \eqref{eq:DAE_error_dynamics_dist}, the LMI in \eqref{eq:H_infinity} can be utilized. 

\vspace{-0.23cm}
\section{Numerical Results and Discussions}\label{sec:numerical_tests}
In this section, we perform numerical simulations to study the applicability and performance of the proposed DAE observers in estimating the trajectories of both algebraic and dynamic states of selected IEEE test networks under various conditions around a certain operating point. All simulations are performed using MATLAB R2020a running on 64-bit Windows 10 with a
%2.5GHz Intel\textsuperscript{R} Core\textsuperscript{TM} i7-6500U CPU
3.4GHz Intel\textsuperscript{R} Core\textsuperscript{TM} i7-6700 CPU
 and 16 GB of RAM, whereas all convex linear programs (LPs) and SDPs are solved through YALMIP \cite{Lofberg2004} optimization interface along with MOSEK \cite{Andersen2000} solver. The dynamical simulations for all DAEs are carried out using MATLAB's ODEs solver \texttt{ode15i}.

\begin{figure}
	\vspace{-0.1cm}
	\centering 
	\subfloat[\label{fig:ssec_B_case9_1}]{\includegraphics[keepaspectratio=true,scale=0.68]{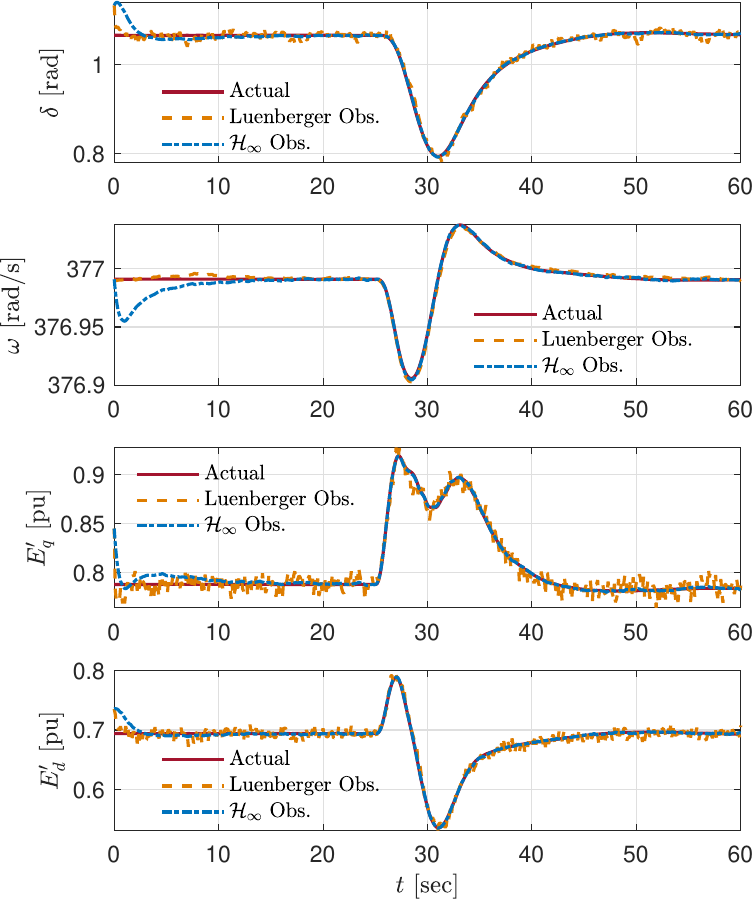}}{}{}\vspace{-0.25cm}
	\subfloat[\label{fig:ssec_B_case9_2}]{\includegraphics[keepaspectratio=true,scale=0.68]{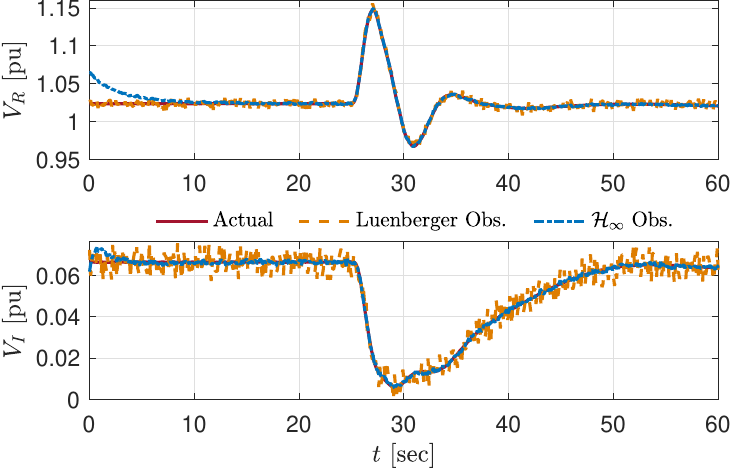}}{}{}\hspace{-0.1cm}
	\vspace{-0.1cm}
	\caption{State estimation results for \textsl{Case-9} with Gaussian noise: \textit{(a)} the internal states of Generator-$2$ and \textit{(b)} complex voltage at Bus-$8$.}
	\label{fig:ssec_B_case9}\vspace{-0.2cm}
\end{figure}

\vspace{-0.23cm}
\subsection{Power Networks Test Cases: Parameters and Setup}\label{ssec:setup}
Herein we consider two power networks of contrasting size: the Western System Coordinating Council
(WSCC) $3$-machine, $9$-bus system (referred to as \textsl{Case-9}) and the New England $9$-machine, $39$-bus system (referred to as \textsl{Case-39}).
{The single-line diagrams for \textsl{Case-9} and \textsl{Case-39} are available in \cite{sauer2017power} and \cite[Appendix A]{Pai1989book}.} The steady-state operating points utilized to construct the matrices in \eqref{eq:SynGenLin}-\eqref{eq:PFLin} are acquired from solving the power flow equations \eqref{eq:GPF} using MATPOWER \cite{MATPOWER2011}. Each synchronous generator parameters appearing in \eqref{eq:SynGen} and \eqref{eq:SynGenStator} are  obtained  from  Power  System Toolbox \cite{sauer2017power} case files \texttt{d3m9bm.m} for \textsl{Case-9} and \texttt{datane.m} for \textsl{Case-39}. The power base for both systems is chosen to be $100$ MVA and synchronous speed $\omega_0 := 2\pi 60\;\mathrm{rad/sec}$. All loads are assumed to be of constant power type. For \textsl{Case-9}, the loads at buses number $5$, $7$, and $9$ are specified as $0.9+j0.3\;\mathrm{pu}$, $1.0+j0.35\;\mathrm{pu}$, and $1.25+j0.5\;\mathrm{pu}$. Since \textsl{Case-39} consists of $21$ loads, they are not specified here---the details can be seen from MATPOWER data file named \texttt{case39.m}.

\begin{figure}
	\vspace{-0.1cm}
	\centering 
	{\includegraphics[keepaspectratio=true,scale=0.68]{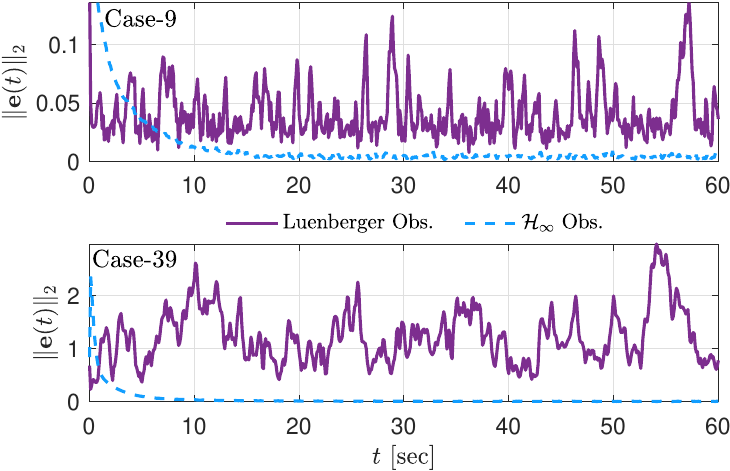}}{}{}
	\vspace{-0.25cm}
	\caption{The comparison of estimation error norm for \textsl{Case-9} and \textsl{Case-39} with Gaussian noise.}
	\label{fig:ssec_B_error}\vspace{-0.1cm}
\end{figure} 

\begin{figure}
	\vspace{-0.05cm}
	\centering 
	\subfloat[\label{fig:ssec_B_case9_case39_1}]{\includegraphics[keepaspectratio=true,scale=0.68]{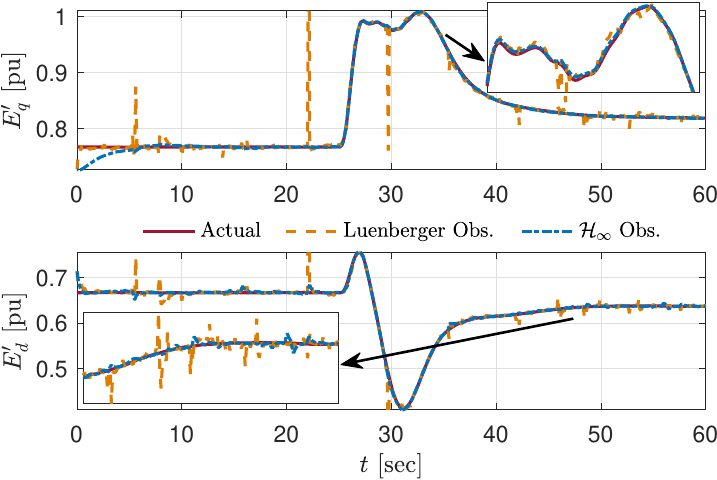}}{}{}\vspace{-0.25cm}
	\subfloat[\label{fig:ssec_B_case9_case39_2}]{\includegraphics[keepaspectratio=true,scale=0.68]{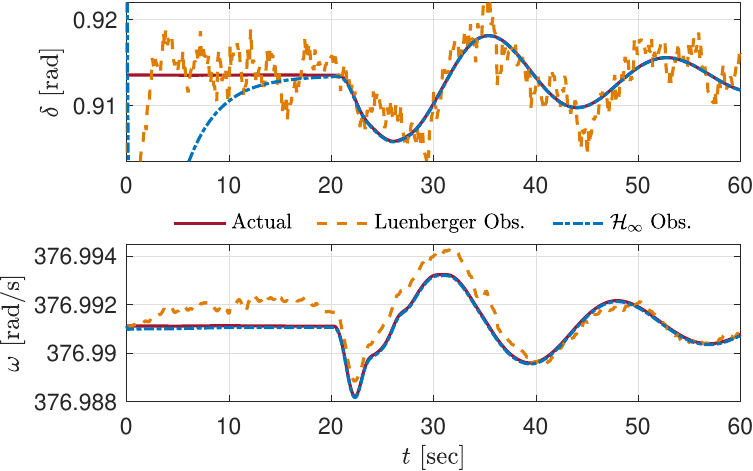}}{}{}\hspace{-0.1cm}
	\vspace{-0.1cm}
	\caption{State estimation results with non-Gaussian measurement noise: \textit{(a)} transient voltage at Generator-$3$ with Cauchy noise (\textsl{Case-9}) and \textit{(b)} rotor angle and speed at Generator-$10$ with Laplace noise (\textsl{Case-39}).}
	\label{fig:ssec_B_case9_case39}\vspace{-0.4cm}
\end{figure}

\vspace{-0.3cm}
\subsection{DSE Under a Three-Phase Fault Contingency}\label{ssec:3_phase_fault}
To examine the performance of the proposed observers in performing DSE under a contingency event, a dynamic response is generated by applying a three-phase fault, where for \textsl{Case-9}, the fault occurs at Bus-$4$ of Line $4$-$5$ at $t = 25\,\mathrm{sec}$ within a $60\,\mathrm{sec}$ simulation window and then cleared at the near and remote ends after $50 \,\mathrm{msec}$ and $200 \,\mathrm{msec}$. 
The three-phase fault for \textsl{Case-39} occurs at Bus-$3$ of Line $3$-$4$ at $t = 20\,\mathrm{sec}$, which is cleared at the near and remote ends after $50 \,\mathrm{msec}$ and $70 \,\mathrm{msec}$. 
It is assumed herein that the input $\m u(t)$ is known to the observers at all times.
We also impose a Gaussian process noise assuming a diagonal covariance matrix which entries are the square of $10$\% of the largest state changes \cite{Zhou2013} as well as a Gaussian measurement noise with variance $0.01^2$ for \textsl{Case-9} and $0.005^2$ for \textsl{Case-39}. Particularly for \textsl{Case-9}, two PMUs are installed with configuration $\mathcal{N}_M = \{4,6\}$, where for \textsl{Case-39}, we follow \cite{Chakrabarti2008} to install $9$ PMUs such that $\mathcal{N}_M = \{2, 6, 10, 19, 20, 22, 23, 25, 29\}$. 
{It is worth noting that these PMU locations render the power networks to be detectable and I-observable---according to \eqref{eq:detectability} and \eqref{eq:I-observability-dae}.} 
The two observers presented in Section \ref{sec:DAE_observer} are implemented: the gain for the standard Luenberger DAE observer is obtained from solving \textbf{P1} while the one for the $\mathcal{H}_{\infty}$ observer is computed from solving \textbf{P2}, from which we get $\kappa = 2.2\times10^{-3}$, $\gamma = 5.78\times10^{-2}$ for \textsl{Case-9} and $\kappa = 1$, $\gamma = 5.24\times10^{-2}$ for \textsl{Case-39}. The initial conditions for observer's states are randomized with $10$\% maximum deviation from the actual states' steady-state values, except for generators' rotor speed, which are set to be equal to the synchronous speed $\omega_0$. 
The disturbance matrices are chosen to be $\m B_w := \bmat{\m I\;\;\m O}\in\mbb{R}^{n\times q}$ and $\m D_w := \bmat{\m O\;\;\m I}\in\mbb{R}^{p\times q}$ where $q = n + p$.  

The results of this numerical study on \textsl{Case-9} are provided in Fig. \ref{fig:ssec_B_case9}.
 Despite Bus-$2$ and Bus-$8$ are not equipped with PMUs, their terminal voltages and Generator-$2$'s internal states can be estimated. It is worth  noting the high noise attenuation provided by the $\mathcal{H}_{\infty}$ DAE observer, compared to the standard Luenberger DAE observer. Similar results are also observed for \textsl{Case-39}, which are not shown here for conciseness. The comparison of estimation error norm is shown in Fig. \ref{fig:ssec_B_error}. It can be seen that the  $\mathcal{H}_{\infty}$ observer provides a superior state tracking---this is indicated by the low estimation error norm.
Motivated by the work in \cite{Wang2018}, we briefly test the DAE observers in handling non-Gaussian measurement noise such as Cauchy and Laplace noise. Following \cite{Nugroho2020DSE}, the Cauchy noise is generated by setting $\m{w}_{mi} = a+ b \cdot \mathrm{tan}\big(\pi(R_2-0.5)\big)$, where $a = 0$, $b = 5\times 10^{-4}$, and $R_2$ is a random number inside $(0,1)$ and implemented on \textsl{Case-9}, 
while the Laplace noise is characterized by the signal $\m{w}_{mi} = m - s \cdot \mathrm{sgn}(R_1)\cdot\mathrm{ln}(1- 2\vert R_1\vert)$,
where $m = 0$, $s = 1\times 10^{-3}$, and $R_1$ is a number randomly chosen inside the set $(-0.5,0.5]$ and implemented on \textsl{Case-39}. The results are presented in Fig. \ref{fig:ssec_B_case9_case39}, from which it can be claimed the superiority of the $\mathcal{H}_{\infty}$ DAE observer over the Luenberger observer.

\setlength{\textfloatsep}{10pt}
\begin{table}[t]
	%	\footnotesize
	\scriptsize
	\vspace{-0.0cm}%\hspace{-0.21cm}
	\centering 
	\caption{The corresponding RMSE \eqref{eq:RMSE} for various configurations of PMU on \textsl{Case-9} with Luenberger and $\mathcal{H}_{\infty}$ DAE observers.}
	\label{tab:PMU_est_qual_case9}
	\vspace{-0.2cm}
	\renewcommand{\arraystretch}{1.5}
	\begin{threeparttable}
\begin{tabular}{c|l|c|c}
		\toprule\hline
	\multirow{2}{*}{Remark} & \multirow{2}{*}{PMU conf. $\mathcal{N}_M$} & \multicolumn{2}{c}{RMSE} \\ \cline{3-4} 
	&                    & Luenberger Obs.          & $\mathcal{H}_{\infty}$ Obs.  \\ \hline
	\multirow{3}{*}{\makecell{$2$ PMUs: \\ $3$-branch bus only} }  &        $\{4,6\}$&   $0.906128$  &   $0.149777$   \\ \cline{2-4} 
	&  $\{4,8\}$&    $0.857766$  &    $0.131873$  \\ \cline{2-4} 
	&  $\{6,8\}$   &  $0.586053$  & $0.128455$ \\ \hline
	\multirow{3}{*}{\makecell{$3$ PMUs: \\ two $3$-branch bus, \\ one load bus} }  &   $\{4,5,8\}$  & $0.906862$ & $0.128968$  \\ \cline{2-4} 
	& $\{4,7,8\}$  &  $0.833687$ & $0.130704$   \\ \cline{2-4} 
	& $\{4,8,9\}$  &  $0.680171$  & $0.118765$  \\ \hline
	\multirow{3}{*}{\makecell{$4$ PMUs: \\ two generator bus, \\ two load bus} }  &  $\{1, 2, 5, 7\}$  & $0.734701$ & $0.141977$ \\ \cline{2-4} 
	& $\{1, 3, 7, 9\}$ & $0.764526$ &  $0.129512$ \\ \cline{2-4} 
%	&                    &            &           \\ \cline{2-4} 
	& $\{2, 3, 5, 9\}$ & $0.686932$  & $0.126877$ \\ \hline
	\multirow{4}{*}{\vspace{0.2cm}\makecell{multiple PMUs: \\ two $3$-branch bus, \\ multiple load bus} }  &  $\{4,5,6\}$ & $0.757261$\tnote{$\dagger$} & $0.153421$\tnote{$\dagger$} \\ \cline{2-4} 
	&  $\{4,5,6,7\}$   &  $0.746314$\tnote{$\dagger$}  &    $0.145797$\tnote{$\dagger$}       \\ \cline{2-4} 
%	&                    &            &           \\ \cline{2-4} 
	& $\{4,5,6,7,9\}$  & $0.644191$\tnote{$\dagger$}  & $0.127354$\tnote{$\dagger$} \\ \hline
	\bottomrule
\end{tabular}
\begin{tablenotes}
	\item[$\dagger$] this value corresponds to the average of RMSE after performing the DSE five times to compensate for noise variability.

\end{tablenotes}
\end{threeparttable}
	\vspace{-0.2cm}
\end{table}
\setlength{\floatsep}{10pt} 

\setlength{\textfloatsep}{10pt}
\begin{table}
	\scriptsize
	\vspace{-0.1cm}%\hspace{-0.21cm}
	\centering 
	\caption{The comparison of RMSE \eqref{eq:RMSE} for DSE with unknown inputs using five different types of observers.}
	\label{tab:DSE_UI}
	\vspace{-0.2cm}
	\renewcommand{\arraystretch}{1.5}
	\begin{threeparttable}
		\begin{tabular}{c|c|c|c|c|c}
				\toprule
			\hline
			\multirow{2}{*}{\textsl{Case}} & \multicolumn{5}{c}{RMSE \& $\gamma$ (for $\mathcal{H}_{\infty}$ Observers only)}  \\ \cline{2-6} 
			& Luenb. Obs. & $\mathcal{H}_{\infty}$ Obs. & PI Luenb.   & S-PI $\mathcal{H}_{\infty}$  & O-PI $\mathcal{H}_{\infty}$ \\ \hline
\multirow{2}{*}{\textsl{9}} & \multirow{2}{*}{$0.9100$} & \multirow{2}{*}{\makecell{$0.5030$\\ $\gamma = 1.444$}} & \multirow{2}{*}{$0.6733$} & \multirow{2}{*}{\makecell{$0.4396$\\ $\gamma = 0.968$}} & \multirow{2}{*}{\makecell{$0.1890$\\ $\gamma = 1.068$}} \\ 
&                    &                   &                   &                   &                   \\ \hline
\multirow{2}{*}{\textsl{39}} & \multirow{2}{*}{$3.6294$} & \multirow{2}{*}{\makecell{$1.1531$\\ $\gamma = 1.318$}} & \multirow{2}{*}{$2.3175$} & \multirow{2}{*}{\makecell{$0.2085$\\ $\gamma = 1.318$}} & \multirow{2}{*}{\makecell{$0.0779$\\ $\gamma = 1.318$}} \\ 
&                    &                   &                   &                   &                   \\ \hline \bottomrule
		\end{tabular}
	\end{threeparttable}
	\vspace{-0.2cm}
\end{table}
\setlength{\floatsep}{10pt}

\begin{figure}
	\vspace{-0.3cm}
	\centering 
	\subfloat[\label{fig:ssec_D_case9_1}]{\includegraphics[keepaspectratio=true,scale=0.68]{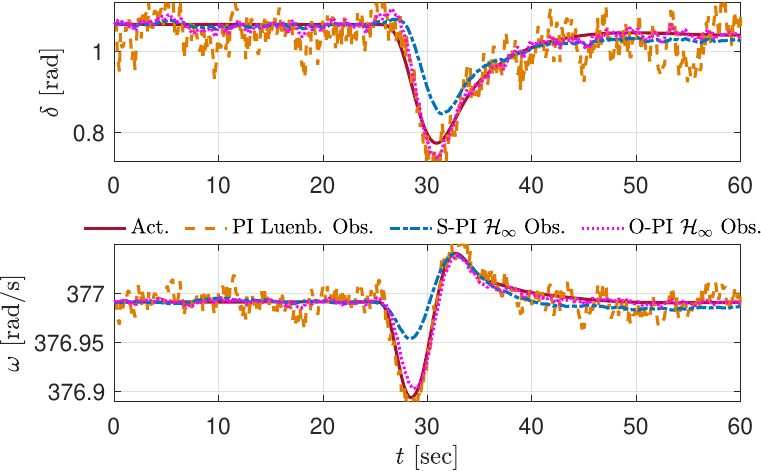}}{}{}\vspace{-0.25cm}
	\subfloat[\label{fig:ssec_D_case9_2}]{\includegraphics[keepaspectratio=true,scale=0.68]{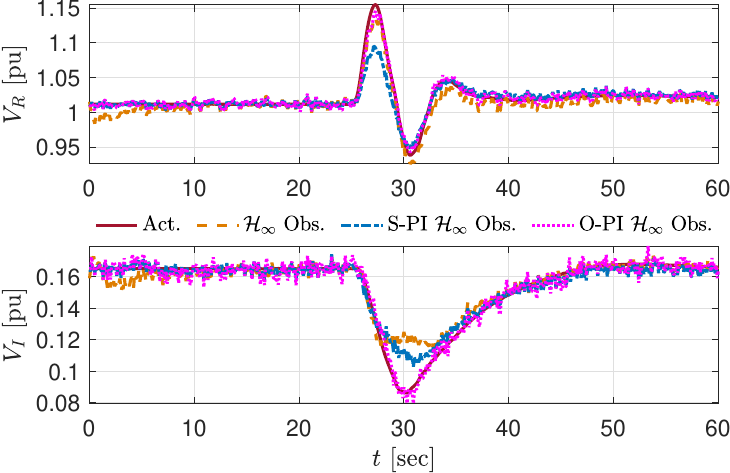}}{}{}\hspace{-0.0cm}
	\vspace{-0.1cm}
	\caption{State estimation results for \textsl{Case-9} with unknown inputs: \textit{(a)} the rotor angle and speed of Generator-$2$ and \textit{(b)} complex voltage at Bus-$2$.}
	\label{fig:ssec_D_case9}\vspace{-0.25cm}
\end{figure}

\vspace{-0.2cm}
\subsection{Effect of PMU's Configuration on Estimation Quality}\label{ssec:PMU_location}
This section studies the influence of varying number of PMU together with different configuration towards the quality of DSE and as such, we limit our study to \textsl{Case-9} only. {The \textit{root-mean-square} error (RMSE) \cite{Nugroho2020DSE,Rouhani2018,Gol2014}, a metric used in the DSE literature, is utilized  to quantify the estimation quality
\begin{align}
	\mathrm{RMSE} &= \sum_{i=1}^{n} \sqrt{\frac{1}{k_f}\sum_{k=0}^{k_f}(e_i[k])^2}, \label{eq:RMSE}
\end{align}  
where $\m e[k]$ is the sampled estimation error and $k_f$ denotes the final time such that $k_f T = 60\,\mathrm{sec}$ where $T = 0.05\;\mathrm{sec}$ is a constant period such that $\m e[k] = \m x(kT) - \hat{\m x}(kT)$.} The results of this study are shown in Tab. \ref{tab:PMU_est_qual_case9}. The first column in Tab. \ref{tab:PMU_est_qual_case9} describes different scenarios pertaining to the number and location of PMUs, whereas the second column shows the PMU configurations. It can be seen that, for the same number of PMUs, both observers yield varying RMSE. The $\mathcal{H}_{\infty}$ DAE observer produces smaller RMSE compared to the Luenberger observer. It is also observed that different placement of PMUs produces different RMSE. There is no clear pattern to indicate that particular PMU locations return better or worse estimation error and/or noise attenuation.
To that end, a further study is required to reveal the relation between PMU locations and estimation quality in a dynamic-algebraic state estimation framework.
Nonetheless, we observe that adding more PMUs generally reduces the RMSE in both observers. This observation is expected since increasing the number of PMUs implies that more states are measured (and therefore, less states are needed to be estimated), thereby decreasing the total estimation error.

\begin{figure}
	\vspace{-0.05cm}
	\centering 
	\subfloat[\label{fig:ssec_D_case39_1}]{\includegraphics[keepaspectratio=true,scale=0.68]{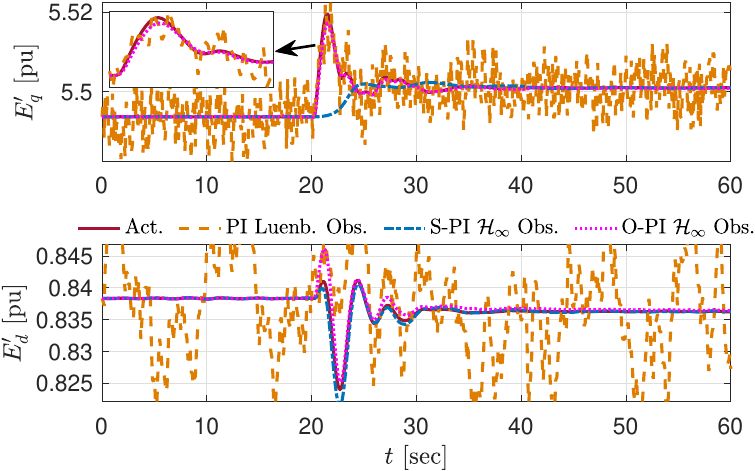}}{}{}\vspace{-0.25cm}
	\subfloat[\label{fig:ssec_D_case39_2}]{\includegraphics[keepaspectratio=true,scale=0.68]{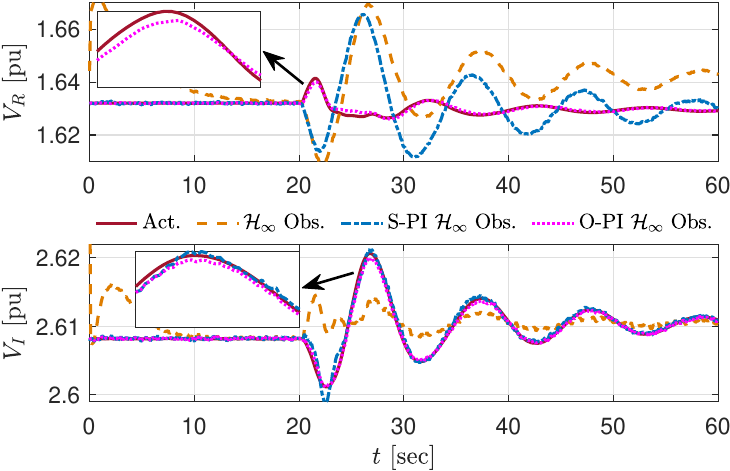}}{}{}\hspace{-0.0cm}
	\vspace{-0.1cm}
	\caption{State estimation results for \textsl{Case-39} with unknown inputs: \textit{(a)} the transient voltage of Generator-$2$ and \textit{(b)} complex voltage at Bus-$8$.}
	\label{fig:ssec_D_case39}\vspace{-0.2cm}
\end{figure}

\vspace{-0.3cm}
\subsection{Estimation Performance with Unmeasurable Inputs}\label{ssec:unmeasurable_inputs}
{As mentioned in Section \ref{sec:DAE_observer_UI}, $T_{Mi}$ and $E_{fdi}$ are difficult to measure practically. To that end, here we consider the case when the actual control inputs are unknown to the observers. Instead, the observers are given static, steady-state control inputs \cite{sauer2017power}.} Note that, due to the three-phase fault, the actual $T_{Mi}$ and $E_{fdi}$ for all $i \in \mathcal{G}$ are fluctuating and settle to a slightly different steady-state value. For this purpose we choose $\m B_{\nu} 
:= \m B_{u}$ such that $\nu = m$. 
The PMU locations and Gaussian noise for both power networks are set to remain the same as in Section \eqref{ssec:3_phase_fault}, with an exception that the variance of Gaussian measurement noise is increased to $0.05^2$ for \textsl{Case-9}. Five different observers are benchmarked:  
%\vspace{-0.2cm}
\begin{itemize}[leftmargin=*]
	\item \textit{Luenb. Obs.:} the Luenberger observer \eqref{eq:DAE_Luenberger_obs} with \textbf{P1}.
	\item \textit{$\mathcal{H}_{\infty}$ Obs.:} the robust $\mathcal{H}_{\infty}$ observer \eqref{eq:DAE_Luenberger_obs} with \textbf{P2}.
	\item \textit{PI Luenb. Obs.:} the PI Luenberger observer \eqref{eq:DAE_base_aug} with \textbf{P1}.
	\item \textit{S-PI $\mathcal{H}_{\infty}$ Obs.:} the robust PI $\mathcal{H}_{\infty}$ observer \eqref{eq:DAE_base_aug} with \textbf{P2}.
	\item \textit{O-PI $\mathcal{H}_{\infty}$ Obs.:} the robust PI $\mathcal{H}_{\infty}$ observer \eqref{eq:DAE_base_aug} with \textbf{P3}.
\end{itemize}

The performance matrices for the robust observers are chosen to be $\m \Gamma := 0.5\m I$ and $\m \Gamma_{\xi} := 0.1\m I$, whereas the constants in \textbf{P3} for the O-PI $\mathcal{H}_{\infty}$ observer are set to be $c_1 = c_2 = 1$. The state estimation results are given in Fig. \ref{fig:ssec_D_case9} for \textsl{Case-9}. Specifically, in Fig. \ref{fig:ssec_D_case9_1}, we show the comparison of three PI observers in estimating some internal states of Generator-$2$. It is clear that the estimation from PI Luenberger observer is the most vulnerable to noise. Strong noise attenuation is demonstrated by the S-PI $\mathcal{H}_{\infty}$ observer. However, it is relatively inaccurate in tracking the actual states trajectory for $25\hspace{-0.00cm}\;\mathrm{sec} \leq\hspace{-0.06cm}t\hspace{-0.06cm}\leq\hspace{-0.06cm} 40\;\mathrm{sec}$. This is presumed to be caused by the assumed steady-state unknown input dynamics \cite{Soffker1995}. The O-PI $\mathcal{H}_{\infty}$ observer provides a sufficient noise attenuation at steady-state (although not as good as the S-PI $\mathcal{H}_{\infty}$ observer) as well as small tracking error when the states are fluctuating. Fig. \ref{fig:ssec_D_case9_2} depicts the estimates of complex voltage at Bus-$2$ with the three $\mathcal{H}_{\infty}$ observers. It is seen that the standard $\mathcal{H}_{\infty}$ observer provides the least accurate estimation compared to the robust PI observers. Despite the fact that O-PI  $\mathcal{H}_{\infty}$ observer is able to properly track the actual states especially for $25\hspace{-0.00cm}\;\mathrm{sec} \leq\hspace{-0.05cm}t\hspace{-0.05cm}\leq\hspace{-0.05cm} 40\;\mathrm{sec}$ with a relatively small error, at steady-state, it is observed from a detailed inspection that its state estimates are experiencing more fluctuations compared to the resulting state estimates from the S-PI $\mathcal{H}_{\infty}$ observer, which is a consequence of higher performance index $\gamma$ from the O-PI $\mathcal{H}_{\infty}$ observer---as seen from Tab. \ref{tab:DSE_UI}. Note that higher $\gamma$ means lower noise attenuation. 
{Similar results are also obtained from \textsl{Case-39}---see Fig. \ref{fig:ssec_D_case39}.}
The summary of RMSE for both cases is presented in 
Tab. \ref{tab:DSE_UI}, where it is seen that the O-PI $\mathcal{H}_{\infty}$ observer has the smallest RMSE. {Thus, it can be concluded that the O-PI $\mathcal{H}_{\infty}$ observer is able to provide superior state estimates, relative to the other observers, when the inputs are unknown.}
% Numerical studies on cases where the input sensors are faulty are provided in \cite{Nugrohotcns2020}.}

%\setlength{\textfloatsep}{10pt}
%\begin{table}
%	%	\footnotesize
%	\scriptsize
%	\vspace{-0.1cm}%\hspace{-0.21cm}
%	\centering 
%	\caption{The comparison of RMSE \eqref{eq:RMSE} for DSE with sensor failures using five different types of observers.}
%	\label{tab:DSE_SF}
%	\vspace{-0.2cm}
%	\renewcommand{\arraystretch}{1.5}
%	\begin{threeparttable}
%		\begin{tabular}{c|c|c|c|c|c}
%			\toprule
%			\hline
%			\multirow{2}{*}{\textsl{Case}} & \multicolumn{5}{c}{RMSE } \\ \cline{2-6} 
%			& Luenb. Obs. & $\mathcal{H}_{\infty}$ Obs. & PI Luenb.   & S-PI $\mathcal{H}_{\infty}$  & O-PI $\mathcal{H}_{\infty}$ \\ \hline
%			\textsl{9} & $0.2362$  & $1.4052$ & $0.1375$  & $0.4559$  & $0.0645$ \\ \hline
%			\textsl{39}  & $2.9481$  & $1.4941$   &  $1.9156$  & $0.1585$   & $0.1365$   \\ \hline
%			\bottomrule
%		\end{tabular}
%	\end{threeparttable}
%	\vspace{-0.1cm}
%\end{table}
%\setlength{\floatsep}{10pt}

\begin{figure}
	\vspace{-0.1cm}
	\centering 
	\subfloat[\label{fig:ssec_F_case9}]{\includegraphics[keepaspectratio=true,scale=0.68]{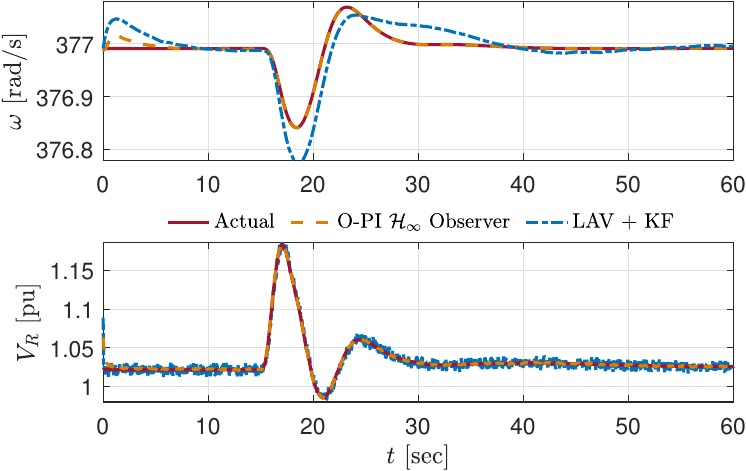}}{}{}\vspace{-0.25cm}
	\subfloat[\label{fig:ssec_G_case9}]{\includegraphics[keepaspectratio=true,scale=0.68]{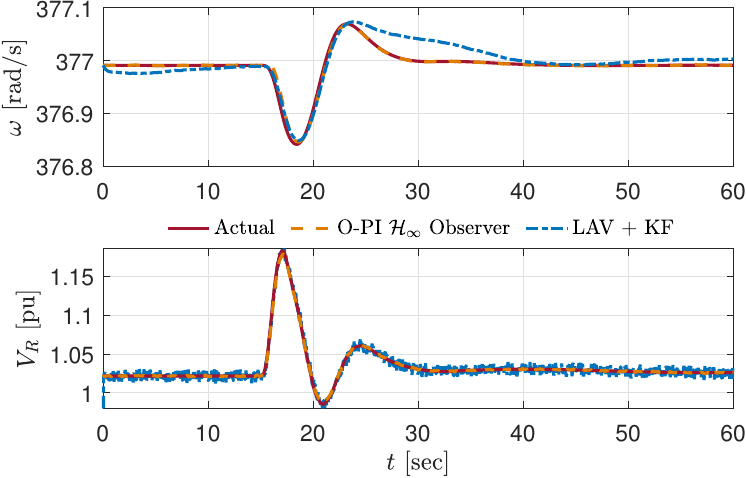}}{}{}\vspace{-0.25cm}
	\subfloat[\label{fig:ssec_FG_case9_error}]{\includegraphics[keepaspectratio=true,scale=0.68]{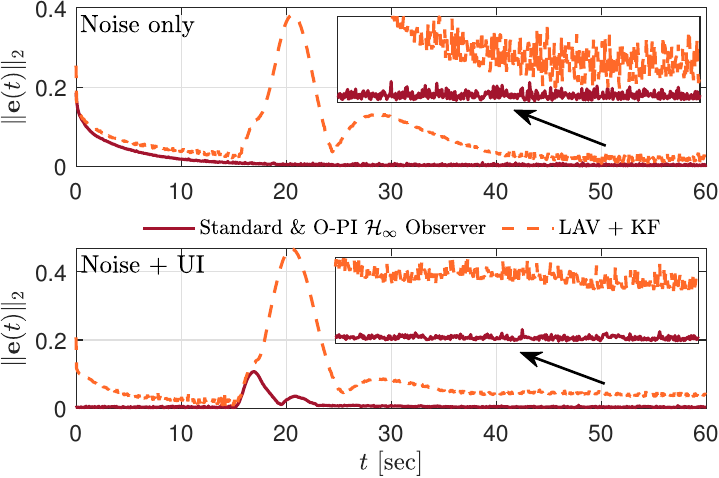}}{}{}\hspace{-0.0cm}
	\vspace{-0.1cm}
	\caption{{The estimation of Generator-$3$ rotor speed and Bus-$3$ real voltage in \textsl{Case-9}: \textit{(a)} the $\mathcal{H}_{\infty}$ observer and LAV-KF with noise only, \textit{(b)} the O-PI $\mathcal{H}_{\infty}$ observer and LAV-KF with noise and unknown inputs (UI), and \textit{(c)} the corresponding estimation error norm.}}
	\label{fig:ssec_FG_case9}\vspace{-0.2cm}
\end{figure}

{
\vspace{-0.2cm}
\subsection{Comparison with A Robust Two-Stage DSE Method}\label{ssec:comparative_study}
In this section we finally compare the proposed DSE framework utilizing the standard and O-PI $\mathcal{H}_{\infty}$ observers with the two-stage approach introduced in \cite{Rouhani2018} to perform DSE on \textsl{Case-9}. In the first stage of the latter approach, a linear phasor estimator is utilized to estimate bus voltages based on measurements provided by the PMUs. The estimated bus voltage values are later used to aid DSE in estimating generators' states in the second stage. The linear phasor estimator implements the {Least Absolute Value} (LAV) method pioneered in \cite{Gol2014} for the purpose of power system's state estimation. It is demonstrated therein that the LAV method can provide better estimates than the weighted least square method in some instances. Due to the two-stage nature, this approach can only be implemented in discrete-time. In each iteration corresponding to a time step $k$, the noisy measurements from PMUs---denoted by $\tilde{\m y}[k]$---are sampled. Afterwards, the following convex LP is solved \cite{Gol2014}
\begin{align*}
	\mathbf{P4}\;\min_{\hat{\m r},\hat{\m v}}\;\; &\sum_{i = 1}^{p} \abs{\hat{r}_i};   \;\; \subjectto \; \tilde{\m y}[k] =\tilde{\m C}\m C_M\hat{\m v} + \hat{\m r},
\end{align*}
where vectors $\hat{\m v}\in\mbb{R}^{2N}$ and $\hat{\m r}\in\mbb{R}^{r}$ denote the estimates of bus voltages and measurement noise, respectively. Using $\hat{\m v}[k]$ and our knowledge on $T_{Mi}$ and $E_{fdi}$ for all $i\in\mc{G}$, the dynamic states can be estimated by implementing a simple discrete-time KF \cite{Kalman1960}.   
}

{In the first instance of this comparative simulation, the presence of Gaussian process and measurement noise are assumed. Three PMUs are installed such that $\mathcal{N}_M = \{4,6,8\}$. The estimators also have the perfect information of $T_{Mi}$ and $E_{fdi}$ at all times. In the two-stage approach (coined as {LAV + KF}), the LAV solves \textbf{P4} in every iteration whereas the KF utilizes the discrete-time dynamics of the power network discretized using the forward Euler method. The time period for this case is chosen to be $0.05\;\mathrm{sec}$. The transient response is generated by introducing a fault at $t = 15\;\mathrm{sec}$. The comparison results are presented in Fig. \ref{fig:ssec_F_case9}. It is apparent that the $\mathcal{H}_{\infty}$ observer provides better estimates than the two-stage approach: the LAV-KF yields less accurate estimates during transient periods, in addition to the noisy estimates. Next, in the second instance, we assume that the estimators do not have access to the actual time-varying values of $T_{Mi}$ and $E_{fdi}$ and instead, they are only fed with their estimated steady-state values. Fig. \ref{fig:ssec_G_case9} shows the comparison of estimation results between the O-PI $\mathcal{H}_{\infty}$ observer with the two-stage approach. The superiority of the standard and O-PI $\mathcal{H}_{\infty}$ observers is also evident from the much smaller estimator error norm---see Fig. \ref{fig:ssec_FG_case9_error}. On the matter of simulation running time, the two-stage approach is also considerably much slower: $644.9\;\mathrm{sec}$ against $15.7\;\mathrm{sec}$ for the case with noise; $629.3\;\mathrm{sec}$ against $14.6\;\mathrm{sec}$ for the case with noise and unknown inputs.}

\vspace{-0.2cm}
\section{Concluding Remarks and Future Work}\label{sec:conclusion}
This paper develops a novel DSE framework on the basis of a detailed, linearized DAE model of power systems---considering a high-order of generator's dynamics, stator's algebraic constraints, generator's complex power, and the network's power balance equations---with PMU-based measurements sensing only bus voltages and line currents. A novel robust $\mathcal{H}_{\infty}$ DAE observer is proposed, which only requires detectability and impulse observability to operate---these conditions are easily satisfied for the proposed power network's model with a few PMUs. 
{
The numerical test results showcase the performance of several observers in estimating generators' internal states and unmeasured bus voltages under various conditions. In particular, it is revealed that the standard and O-PI robust $\mathcal{H}_{\infty}$ observers are superior compared to the Luenberger observers and the two-stage approach that is based on LAV and KF in performing DSE in the presence of noise and unknown inputs.}   

{
 The utilization of a linear DAE representation of transmission power networks inevitably becomes the major limitation in this approach. On that regard, extending the proposed approach based on a nonlinear, DAE representation of power networks is a worthy direction for future research. We also plan to study the effectiveness of the proposed approach in managing sensor failures from PMUs as well as potential cyber-attacks.}

\vspace{-0.2cm}

\bibliographystyle{IEEEtran}	\bibliography{bib_file}

\begin{thebibliography}{10}
\providecommand{\url}[1]{#1}
\csname url@samestyle\endcsname
\providecommand{\newblock}{\relax}
\providecommand{\bibinfo}[2]{#2}
\providecommand{\BIBentrySTDinterwordspacing}{\spaceskip=0pt\relax}
\providecommand{\BIBentryALTinterwordstretchfactor}{4}
\providecommand{\BIBentryALTinterwordspacing}{\spaceskip=\fontdimen2\font plus
\BIBentryALTinterwordstretchfactor\fontdimen3\font minus
  \fontdimen4\font\relax}
\providecommand{\BIBforeignlanguage}[2]{{%
\expandafter\ifx\csname l@#1\endcsname\relax
\typeout{** WARNING: IEEEtran.bst: No hyphenation pattern has been}%
\typeout{** loaded for the language `#1'. Using the pattern for}%
\typeout{** the default language instead.}%
\else
\language=\csname l@#1\endcsname
\fi
#2}}
\providecommand{\BIBdecl}{\relax}
\BIBdecl

\bibitem{Chakrabortty2013}
A.~{Chakrabortty} and P.~P. {Khargonekar}, ``Introduction to wide-area control
  of power systems,'' in \emph{2013 American Control Conference}, 2013, pp.
  6758--6770.

\bibitem{Zhao2019PSDSE}
J.~{Zhao}, A.~{Gómez-Expósito}, M.~{Netto}, L.~{Mili}, A.~{Abur},
  V.~{Terzija}, I.~{Kamwa}, B.~{Pal}, A.~K. {Singh}, J.~{Qi}, Z.~{Huang}, and
  A.~P.~S. {Meliopoulos}, ``Power system dynamic state estimation: Motivations,
  definitions, methodologies, and future work,'' \emph{IEEE Transactions on
  Power Systems}, vol.~34, no.~4, pp. 3188--3198, 2019.

\bibitem{Qi2018Access}
J.~{Qi}, A.~F. {Taha}, and J.~{Wang}, ``Comparing kalman filters and observers
  for power system dynamic state estimation with model uncertainty and
  malicious cyber attacks,'' \emph{IEEE Access}, vol.~6, pp. 77\,155--77\,168,
  2018.

\bibitem{Nugroho2020DSE}
S.~A. {Nugroho}, A.~F. {Taha}, and J.~{Qi}, ``Robust dynamic state estimation
  of synchronous machines with asymptotic state estimation error performance
  guarantees,'' \emph{IEEE Transactions on Power Systems}, vol.~35, no.~3, pp.
  1923--1935, 2020.

\bibitem{Zhao2017EKF}
J.~{Zhao}, M.~{Netto}, and L.~{Mili}, ``A robust iterated extended kalman
  filter for power system dynamic state estimation,'' \emph{IEEE Transactions
  on Power Systems}, vol.~32, no.~4, pp. 3205--3216, 2017.

\bibitem{Ghahremani2016}
E.~{Ghahremani} and I.~{Kamwa}, ``Local and wide-area pmu-based decentralized
  dynamic state estimation in multi-machine power systems,'' \emph{IEEE
  Transactions on Power Systems}, vol.~31, no.~1, pp. 547--562, 2016.

\bibitem{Qi2018UKF}
J.~{Qi}, K.~{Sun}, J.~{Wang}, and H.~{Liu}, ``Dynamic state estimation for
  multi-machine power system by unscented kalman filter with enhanced numerical
  stability,'' \emph{IEEE Transactions on Smart Grid}, vol.~9, no.~2, pp.
  1184--1196, 2018.

\bibitem{Gross2016}
T.~B. {Gross}, S.~{Trenn}, and A.~{Wirsen}, ``Solvability and stability of a
  power system dae model,'' \emph{Systems \& Control Letters}, vol.~97, pp. 12
  -- 17, 2016.

\bibitem{sauer2017power}
P.~Sauer, M.~Pai, and J.~Chow, \emph{Power System Dynamics and Stability: With
  Synchrophasor Measurement and Power System Toolbox}, ser. Wiley - IEEE.\hskip
  1em plus 0.5em minus 0.4em\relax Wiley, 2017.

\bibitem{gomez2011use}
A.~Gomez-Exposito, A.~Abur, P.~Rousseaux, A.~de~la Villa~Jaen, and
  C.~Gomez-Quiles, ``On the use of pmus in power system state estimation,''
  \emph{Proceedings of the 17th PSCC}, 2011.

\bibitem{Risbud2016}
P.~{Risbud}, N.~{Gatsis}, and A.~{Taha}, ``Assessing power system state
  estimation accuracy with gps-spoofed pmu measurements,'' in \emph{2016 IEEE
  Power Energy Society Innovative Smart Grid Technologies Conference (ISGT)},
  2016, pp. 1--5.

\bibitem{Sarri2016}
S.~{Sarri}, L.~{Zanni}, M.~{Popovic}, J.~{Le Boudec}, and M.~{Paolone},
  ``Performance assessment of linear state estimators using synchrophasor
  measurements,'' \emph{IEEE Transactions on Instrumentation and Measurement},
  vol.~65, no.~3, pp. 535--548, 2016.

\bibitem{Risbud2020}
P.~{Risbud}, N.~{Gatsis}, and A.~{Taha}, ``Multi-period power system state
  estimation with pmus under gps spoofing attacks,'' \emph{Journal of Modern
  Power Systems and Clean Energy}, vol.~8, no.~4, pp. 597--606, 2020.

\bibitem{Rouhani2018}
A.~Rouhani and A.~Abur, ``Linear phasor estimator assisted dynamic state
  estimation,'' \emph{IEEE Transactions on Smart Grid}, vol.~9, no.~1, pp.
  211--219, 2018.

\bibitem{Abur2015}
A.~Abur, ``Observability and dynamic state estimation,'' in \emph{2015 IEEE
  Power Energy Society General Meeting}, 2015, pp. 1--5.

\bibitem{Zhang2014}
J.~Zhang, G.~Welch, G.~Bishop, and Z.~Huang, ``A two-stage kalman filter
  approach for robust and real-time power system state estimation,'' \emph{IEEE
  Transactions on Sustainable Energy}, vol.~5, no.~2, pp. 629--636, 2014.

\bibitem{Dafis2002}
C.~Dafis and C.~Nwankpa, ``A nonlinear observability formulation for power
  systems incorporating generator dynamics,'' in \emph{2002 IEEE International
  Symposium on Circuits and Systems. Proceedings (Cat. No.02CH37353)}, vol.~5,
  2002, pp. V--V.

\bibitem{Rinaldi2017a}
G.~{Rinaldi}, P.~P. {Menon}, C.~{Edwards}, and A.~{Ferrara}, ``Distributed
  observers for state estimation in power grids,'' in \emph{2017 American
  Control Conference (ACC)}, 2017, pp. 5824--5829.

\bibitem{Rinaldi2017b}
G.~{Rinaldi} and A.~{Ferrara}, ``Higher order sliding mode observers and
  nonlinear algebraic estimators for state tracking in power networks,'' in
  \emph{2017 IEEE 56th Annual Conference on Decision and Control (CDC)}, 2017,
  pp. 6033--6038.

\bibitem{Rinaldi2020}
G.~{Rinaldi}, P.~P. {Menon}, C.~{Edwards}, and A.~{Ferrara}, ``Design and
  validation of a distributed observer-based estimation scheme for power
  grids,'' \emph{IEEE Transactions on Control Systems Technology}, vol.~28,
  no.~2, pp. 680--687, 2020.

\bibitem{Taha2019TCNS}
A.~F. Taha, M.~Bazrafshan, S.~A. Nugroho, N.~Gatsis, and J.~Qi, ``Robust
  control for renewable-integrated power networks considering input bound
  constraints and worst case uncertainty measure,'' \emph{IEEE Transactions on
  Control of Network Systems}, vol.~6, no.~3, pp. 1210--1222, 2019.

\bibitem{LorenzMeyer2020}
N.~Lorenz-Meyer, A.~Bobtsov, R.~Ortega, N.~Nikolaev, and J.~Schiffer,
  ``{PMU}-based decentralised mixed algebraic and dynamic state observation in
  multi-machine power systems,'' \emph{{IET} Generation, Transmission {\&}
  Distribution}, vol.~14, no.~25, pp. 6267--6275, Dec. 2020.

\bibitem{astolfi_nonlinear_2008}
A.~Astolfi, D.~Karagiannis, and R.~Ortega, \emph{Nonlinear and adaptive control
  with applications}, ser. Communications and control engineering.\hskip 1em
  plus 0.5em minus 0.4em\relax London: Springer, 2008, oCLC: ocn166315452.

\bibitem{Aranovskiy2017}
S.~Aranovskiy, A.~Bobtsov, R.~Ortega, and A.~Pyrkin, ``Performance enhancement
  of parameter estimators via dynamic regressor extension and mixing*,''
  \emph{IEEE Transactions on Automatic Control}, vol.~62, no.~7, pp.
  3546--3550, 2017.

\bibitem{Emami2015PF}
K.~{Emami}, T.~{Fernando}, H.~H. {Iu}, H.~{Trinh}, and K.~P. {Wong}, ``Particle
  filter approach to dynamic state estimation of generators in power systems,''
  \emph{IEEE Transactions on Power Systems}, vol.~30, no.~5, pp. 2665--2675,
  2015.

\bibitem{Darouach1995}
M.~Darouach and M.~Boutayeb, ``Design of observers for descriptor systems,''
  \emph{IEEE Transactions on Automatic Control}, vol.~40, no.~7, pp.
  1323--1327, 1995.

\bibitem{Hou1999}
M.~Hou and P.~Muller, ``Observer design for descriptor systems,'' \emph{IEEE
  Transactions on Automatic Control}, vol.~44, no.~1, pp. 164--169, 1999.

\bibitem{darouach2009}
M.~Darouach, ``{H-infinity unbiased filtering for linear descriptor systems via
  LMI},'' \emph{{IEEE Transactions on Automatic Control}}, vol.~54, no.~8, pp.
  1966--1972, Aug. 2009.

\bibitem{Xu2003}
S.~Xu, J.~Lam, and Y.~Zou, ``$\mathcal{H}_{\infty}$ filtering for singular
  systems,'' \emph{IEEE Transactions on Automatic Control}, vol.~48, no.~12,
  pp. 2217--2222, 2003.

\bibitem{XU200748}
S.~Xu and J.~Lam, ``Reduced-order $\mathcal{H}_{\infty}$ filtering for singular
  systems,'' \emph{Systems \& Control Letters}, vol.~56, no.~1, pp. 48--57,
  2007.

\bibitem{Nugroho2021CCTA}
S.~A. Nugroho, A.~F. Taha, N.~Gatsis, and J.~Zhao, ``On the simultaneous
  estimation of dynamic and algebraic states in power networks via state
  observer,'' in \emph{2021 IEEE Conference on Control Technology and
  Applications (CCTA)}, 2021, pp. 309--314.

\bibitem{Kekatos2012PMU}
V.~{Kekatos}, G.~B. {Giannakis}, and B.~{Wollenberg}, ``Optimal placement of
  phasor measurement units via convex relaxation,'' \emph{IEEE Transactions on
  Power Systems}, vol.~27, no.~3, pp. 1521--1530, 2012.

\bibitem{Zhou2006PMU}
M.~{Zhou}, V.~A. {Centeno}, J.~S. {Thorp}, and A.~G. {Phadke}, ``An alternative
  for including phasor measurements in state estimators,'' \emph{IEEE
  Transactions on Power Systems}, vol.~21, no.~4, pp. 1930--1937, 2006.

\bibitem{MATPOWER2011}
R.~D. {Zimmerman}, C.~E. {Murillo-Sánchez}, and R.~J. {Thomas}, ``Matpower:
  Steady-state operations, planning, and analysis tools for power systems
  research and education,'' \emph{IEEE Transactions on Power Systems}, vol.~26,
  no.~1, pp. 12--19, 2011.

\bibitem{duan2010analysis}
G.-R. Duan, \emph{Analysis and design of descriptor linear systems}.\hskip 1em
  plus 0.5em minus 0.4em\relax Springer Science \& Business Media, 2010,
  vol.~23.

\bibitem{Gross2014}
T.~B. {Gross}, S.~{Trenn}, and A.~{Wirsen}, ``Topological solvability and index
  characterizations for a common dae power system model,'' in \emph{2014 IEEE
  Conference on Control Applications (CCA)}, 2014, pp. 9--14.

\bibitem{Gupta2014}
M.~K. {Gupta}, N.~K. {Tomar}, and S.~{Bhaumik}, ``Detectability and observer
  design for linear descriptor systems,'' in \emph{22nd Mediterranean
  Conference on Control and Automation}, 2014, pp. 1094--1098.

\bibitem{Luenberger1971}
D.~{Luenberger}, ``An introduction to observers,'' \emph{IEEE Transactions on
  Automatic Control}, vol.~16, no.~6, pp. 596--602, 1971.

\bibitem{xu2006robust}
S.~Xu and J.~Lam, \emph{Robust Control and Filtering of Singular Systems}, ser.
  Lecture Notes in Control and Information Sciences.\hskip 1em plus 0.5em minus
  0.4em\relax Springer Berlin Heidelberg, 2006.

\bibitem{Dai1989}
L.~Dai, \emph{Singular Control Systems}.\hskip 1em plus 0.5em minus 0.4em\relax
  Berlin, Heidelberg: Springer-Verlag, 1989.

\bibitem{Ghahremani2011}
E.~{Ghahremani} and I.~{Kamwa}, ``Dynamic state estimation in power system by
  applying the extended kalman filter with unknown inputs to phasor
  measurements,'' \emph{IEEE Transactions on Power Systems}, vol.~26, no.~4,
  pp. 2556--2566, 2011.

\bibitem{Lee2020}
Y.~{Lee}, S.~H. {Kim}, G.~{Lee}, and Y.~{Shin}, ``Dynamic state estimation of
  generator using pmu data with unknown inputs,'' in \emph{2020 IEEE 29th
  International Symposium on Industrial Electronics (ISIE)}, 2020, pp.
  839--844.

\bibitem{Anagnostou2018}
G.~{Anagnostou} and B.~C. {Pal}, ``Derivative-free kalman filtering based
  approaches to dynamic state estimation for power systems with unknown
  inputs,'' \emph{IEEE Transactions on Power Systems}, vol.~33, no.~1, pp.
  116--130, 2018.

\bibitem{Zhou2013}
N.~{Zhou}, D.~{Meng}, and S.~{Lu}, ``Estimation of the dynamic states of
  synchronous machines using an extended particle filter,'' \emph{IEEE
  Transactions on Power Systems}, vol.~28, no.~4, pp. 4152--4161, 2013.

\bibitem{BAKHSHANDE2015}
F.~Bakhshande and D.~Söffker, ``Proportional-integral-observer: A brief survey
  with special attention to the actual methods using acc benchmark,''
  \emph{IFAC-PapersOnLine}, vol.~48, no.~1, pp. 532 -- 537, 2015, 8th Vienna
  International Conferenceon Mathematical Modelling.

\bibitem{Soffker1995}
D.~Söffker, T.-J. Yu, and P.~C. M{\"u}ller, ``State estimation of dynamical
  systems with nonlinearities by using proportional-integral observer,''
  \emph{International Journal of Systems Science}, vol.~26, no.~9, pp.
  1571--1582, 1995.

\bibitem{Lofberg2004}
J.~L{\"{o}}fberg, ``Yalmip : A toolbox for modeling and optimization in
  matlab,'' in \emph{In Proceedings of the CACSD Conference}, Taipei, Taiwan,
  2004.

\bibitem{Andersen2000}
E.~D. Andersen and K.~D. Andersen, \emph{The Mosek Interior Point Optimizer for
  Linear Programming: An Implementation of the Homogeneous Algorithm}.\hskip
  1em plus 0.5em minus 0.4em\relax Boston, MA: Springer US, 2000, pp. 197--232.

\bibitem{Pai1989book}
M.~Pai, \emph{Energy Function Analysis for Power System Stability}, ser. Power
  Electronics and Power Systems.\hskip 1em plus 0.5em minus 0.4em\relax
  Springer US, 1989.

\bibitem{Chakrabarti2008}
S.~{Chakrabarti} and E.~{Kyriakides}, ``Optimal placement of phasor measurement
  units for power system observability,'' \emph{IEEE Transactions on Power
  Systems}, vol.~23, no.~3, pp. 1433--1440, 2008.

\bibitem{Wang2018}
S.~{Wang}, J.~{Zhao}, Z.~{Huang}, and R.~{Diao}, ``Assessing gaussian
  assumption of pmu measurement error using field data,'' \emph{IEEE
  Transactions on Power Delivery}, vol.~33, no.~6, pp. 3233--3236, 2018.

\bibitem{Gol2014}
M.~Göl and A.~Abur, ``Lav based robust state estimation for systems measured
  by pmus,'' \emph{IEEE Transactions on Smart Grid}, vol.~5, no.~4, pp.
  1808--1814, 2014.

\bibitem{Kalman1960}
R.~E. Kalman, ``A new approach to linear filtering and prediction problems,''
  \emph{Transactions of the ASME--Journal of Basic Engineering}, vol.~82, no.
  Series D, pp. 35--45, 1960.

\end{thebibliography}

\appendices

\vspace{-0.2cm}
\section{The Description of Matrices in \eqref{eq:SynGenLin}, \eqref{eq:SynGenStatorLin}, and  \eqref{eq:PFLin}}\label{appdx:A}
The matrices in generator's linear dynamics \eqref{eq:SynGenLin} can be written as $\m E_D = \mathrm{blkdiag}\left(\{\m E_{Di}\}_{i\in \mc{G}}\right)$, $\m A_D = \mathrm{blkdiag}\left(\{\m A_{Di}\}_{i\in \mc{G}}\right)$, $\m D_D = \mathrm{blkdiag}\left(\{\m D_{Di}\}_{i\in \mc{G}}\right)$, and $\m B_D = \mathrm{blkdiag}\left(\{\m B_{Di}\}_{i\in \mc{G}}\right)$, 	where the matrices for each $i$ are described as follows
%\begin{subequations}\label{eq:SynGenLinMat}
%	\begin{align*}
%	\m E_{Di}\Delta \dot{\tilde{\m x}}_i (t) = \m A_{Di}\Delta \tilde{\m x}_i (t) + \m D_{Di} \Delta {\m i}_{gi}(t) + \m B_{Di} \Delta \tilde{\m u}_i(t),%\label{eq:SynGenLinMat1}
%	\end{align*}
\begin{align*}
	\m E_{Di} &= \mathrm{Diag}(1,M_{i},T'_{d0i},T'_{q0i}),\\
	\m A_{Di} &= \bmat{0&1&0&0\\0 &-{D_i}&-{i^0_{qi}}&-{i^0_{di}}\\0&0&-1&0\\0&0&0&-1}, \m B_{Di}^\top= \bmat{0&1&0&0\\0&0&1&0},\\
	\m D_{Di} &= \bmat{0&0\\{i^0_{qi}(x'_{di}-x'_{qi})-e^{\prime 0}_{di}}&{i^0_{di}(x'_{di}-x'_{qi})-e^{\prime 0}_{qi}}\\-{\left(x_{di} - x'_{di} \right)}&0\\0&{x_{qi} - x'_{qi} }},
\end{align*}
%\end{subequations}
where the corresponding matrices in stator's algebraic equations \eqref{eq:SynGenStatorLin} can be constructed as $\m A_A = \mathrm{blkdiag}\left(\{\m A_{Ai}\}_{i\in \mc{G}}\right)$, $\m D_A = \mathrm{blkdiag}\left(\{\m D_{Ai}\}_{i\in \mc{G}}\right)$, and $\m G_A = \mathrm{blkdiag}\left(\{\m G_{Ai}\}_{i\in \mc{G}}\right)$. The matrices for each $i$ are given as
%	\begin{subequations}\label{eq:SynGenStatorLinMat}
\begin{align*}
	\m A_{Ai} &= \bmat{-v_{Ri}^0\cos\delta_i^0-v_{Ii}^0\sin\delta_i^0&0&0&1\\ v_{Ri}^0\sin\delta_i^0-v_{Ii}^0\cos\delta_i^0&0&1&0},\\
	\m D_{Ai} &= \bmat{-R_{si}&x'_{qi}\\-x'_{di}&-R_{si}}, 
	\m G_{Ai} = \bmat{ -\sin\delta_i^0&\cos\delta_i^0\\ -\cos\delta_i^0&-\sin\delta_i^0}.
\end{align*}
%\end{subequations}
The matrices in \eqref{eq:PFLin} are obtained from linearizing \eqref{eq:GPF} for all generator and load buses. The details on these matrices are not provided here for brevity, as similar procedures can be followed from \cite[Section 8.2]{sauer2017power} to compute them.
\vspace{-0.1cm}

\vspace{-0.2cm}
\section{Proof of Theorem \ref{thm:regular}}\label{appdx:B}
Let $\m M,\;\m N\in\mbb{R}^{n\times n}$ be two nonsingular matrices such that 
$\m M = \Blkdiag(\m E_D^{-1},\m I)$ and $\m N = \m I$. By defining $\tilde{\m E}$, $\tilde{\m A}$, and $\tilde{\m B}_u$ respectively as $\tilde{\m E} := \m M\m E\m N$, $\tilde{\m A} := \m M\m A\m N$, and $\tilde{\m B}_u := \m M\m B_u\m N$, the DAE \eqref{eq:DAE_base}
is a restricted system equivalent to 
\begin{align}
	\tilde{\m E}\dot{\m x}(t) = \tilde{\m A} {\m x}(t) + \tilde{\m B}_u \m u(t), \;\quad\m y(t) = \m C{\m x}(t), \label{eq:DAE_base_rse} 
\end{align} where the matrices $\tilde{\m E}$, $\tilde{\m A}$, and $\tilde{\m B}_u$ are detailed as
\begin{align*}
	\tilde{\m E} = \bmat{\,\m I\;\;\,\m O \\ \m O \;\;\m O},\, \tilde{\m A} = \bmat{\m E_D^{-1}\m A_1 \;\; \m E_D^{-1}\m A_2 \\ \m A_3\;\;\;\;\;\;\; \m A_4},\, \tilde{\m B}_u = \bmat{\m E_D^{-1}\m B_D \\ \m O}. 
\end{align*}
\textcolor{black}{Since $\m A_4$ is nonsingular, then according to \cite[Lemma 2.3]{xu2006robust}, the transformed DAE \eqref{eq:DAE_base_rse} is impulse-free, implying that the DAE is of index one and regular \cite{Dai1989}.} Nonetheless, as \eqref{eq:DAE_base_rse} is a restricted system equivalent to DAE \eqref{eq:DAE_base}, the considered power networks model is also impulse-free, of index one, and regular. This completes the proof.
\newqed
\vspace{-0.1cm}

\vspace{-0.2cm}
\section{Proof of Theorem \ref{thm:I-observability}}\label{appdx:C}
Substituting the matrices $\m E$, $\m A$, and $\m C$ in DAE \eqref{eq:DAE_base} to \eqref{eq:I-observability}, then due to $\m E_D$ being a full rank matrix, we obtain
\begin{align*}
	\rank\hspace{-0.05cm}\left(\hspace{-0.05cm}\bmat{\m E_D & \m A_1 & \m A_2 \\ \m O & \m A_3 & \m A_4 \\ \m O & \m E_D & \m O \\ \m O & \m O & \tilde{\m C}\m C_M}\hspace{-0.05cm}\right) - \rank(\m E_D) &= n \\
	\Leftrightarrow \rank(\m E_D) + \rank\hspace{-0.05cm}\left(\hspace{-0.05cm}\bmat{\;\m A_3 \;\;\;\;\, \m A_4 \;\;\;\\ \m E_D \;\;\;\;\;\m O\;\;\;\; \\ \;\m O \;\;\;\;\tilde{\m C}\m C_M}\hspace{-0.05cm}\right) - \rank(\m E_D) &= n \\
	\Leftrightarrow  \rank\hspace{-0.05cm}\left(\hspace{-0.05cm}\bmat{\m A_4^{-1}&\m O & \m O \\ \m O & \m E_D^{-1} & \m O \\ \m O & \m O & \m I}\hspace{-0.1cm}\bmat{\;\m A_3 \;\;\;\;\, \m A_4 \;\;\;\\ \m E_D \;\;\;\;\;\m O\;\;\;\; \\ \;\m O \;\;\;\;\tilde{\m C}\m C_M}\hspace{-0.05cm}\right) &= n \\
	%\Leftrightarrow  \rank\hspace{-0.05cm}\left(\hspace{-0.05cm}\bmat{\m A_4^{-1}\m A_3 \;\;\;\; \m I\;\;\;\;\;\\ \;\m I \;\;\;\;\;\;\;\;\,\m O \;\;\\\;\; \m O \;\;\;\;\;\;\tilde{\m C}\m C_M}\hspace{-0.1cm}\bmat{\m I & \m O \\ \m O & \m C_M^\top}\hspace{-0.05cm}\right) &= n,
	\Leftrightarrow  \rank\hspace{-0.05cm}\left(\hspace{-0.05cm}\bmat{\m A_4^{-1}\m A_3 & \m I\\ \m I &\m O \\\m O &\tilde{\m C}\m C_M}\hspace{-0.1cm}\bmat{\m I & \m O \\ \m O & \m C_M^\top}\hspace{-0.05cm}\right) &= n,
	%\\
	%\Leftrightarrow\rank\hspace{-0.05cm}\left(\hspace{-0.05cm}\bmat{\m A_4^{-1}\m A_3 \;\; \m C_M^\top \\ \;\;\;\,\m I \;\;\;\;\;\;\;\,\m O \;\;\;\\\ \;\;\m O \;\;\;\;\;\;\;\tilde{\m C}\;\;\;}\hspace{-0.05cm}\right) \hspace{-0.1cm}&= \hspace{-0.00cm}n.
\end{align*}
which is equivalent to \eqref{eq:I-observability-dae}, thereby completing the proof.
\newqed
\vspace{-0.1cm}

\vspace{-0.2cm}
\section{Proof of Theorem \ref{thm:admissibility_error_dyn}}\label{appdx:D}
By substituting $\tilde{\m E} = \m E$ and $\tilde{\m A} = \m A -\m L \m C$ to the matrix inequality \eqref{eq:admissibility} in Lemma \ref{lem:admissibility}, one can immediately obtain 
\begin{align*}
	\m A ^{\top} \m X \m E + \m E^\top\m X\m A + &\m A ^{\top} \m E^{\perp\top} \m Y + \m Y^\top\m E^{\perp}\m A \nonumber \\
	&\;- \m C^{\top}\m L^{\top}\m P -\m P^\top\m L \m C \prec 0. %\label{eq:admissibility_LMI_proof}
\end{align*}
By defining a new matrix $\m W\in\mbb{R}^{n\times p}$ such that $\m W := \m L^{\top}\m P$, the LMI in \eqref{eq:admissibility_LMI} is established. This ends the proof. 
\newqed
\vspace{-0.1cm}

\vspace{-0.2cm}
\section{Proof of Theorem \ref{thm:robust_observer}}\label{appdx:E}
Suppose that the LMI \eqref{eq:H_infinity} holds. Then the $(1,1)$ block matrix in \eqref{eq:H_infinity} implies that the LMI \eqref{eq:admissibility_LMI} is satisfied, thereby indicating that the undisturbed error dynamics in \eqref{eq:DAE_error_dynamics_dist} is stable. Next, consider a Lyapunov function candidate $V:\mbb{R}^n\rightarrow \mbb{R}_+$ where $V(\m e) := \m e(t)^\top \m E^\top \m P \m e(t)$ with $\m P \in \mbb{R}^{n\times n}$ nonsingular and $\m E^\top \m P = \m P^\top \m E \succeq 0$. Then, the time derivative of $V(\cdot)$ along its trajectory can be written as follows
\begin{align}
	\dot{V}(\m e) &= \m e(t)^\top\hspace{-0.05cm}\left(\m A^\top\hspace{-0.05cm}\m P + \m P^\top\hspace{-0.05cm}\m A - \m C^{\top}\hspace{-0.05cm}\m L^{\top}\hspace{-0.05cm}\m P -\m P^\top\hspace{-0.05cm}\m L \m C\hspace{0.05cm} \right)\hspace{-0.05cm}\m e(t) \nonumber \\
	&\quad+\m w(t)^\top\hspace{-0.1cm}\left(\m B_w^\top\hspace{-0.00cm}\m P - \m D_w^\top\hspace{-0.00cm}\m L^\top\hspace{-0.00cm} \m P\right)\hspace{-0.05cm}\m e(t) \nonumber\\
	&\quad+\m e(t)^\top\hspace{-0.1cm}\left(\m P^\top\hspace{-0.00cm}\m B_w - \m P^\top\hspace{-0.00cm}\m L\m D_w\right)\hspace{-0.05cm}\m w(t).\label{eq:robust_observer_proof_1}
\end{align}
%For case \textit{(a)}, since $\m w(t) = \m 0$ for all $t \geq 0$, then from the first block of \eqref{eq:H_infinity}, it implies the LMI \eqref{eq:admissibility_LMI}, which shows that 
For any bounded disturbance $\m w(t)$, then integrating the condition $\dot{V}(\m e) + \m {\epsilon}^\top \m {\epsilon} - \gamma \m w^\top\m w < 0$ from $0$ to $t$ reads to
\begin{align}
	\int_{0}^{t}\dot{V}(\m e(\tau)) d\tau < - \int_{0}^{t}\m {\epsilon}(\tau)^\top \m {\epsilon}(\tau)d\tau + \int_{0}^{t}\gamma \m w(\tau)^\top\m w(\tau) d\tau \nonumber \\
	\Leftrightarrow V(e(t)) - V(e_0) < - \int_{0}^{t}\norm{\m \epsilon(\tau)}^2_2 d\tau + \int_{0}^{t}\hspace{-0.05cm}\gamma \norm{\m w(\tau)}^2_2 d\tau.\nonumber
\end{align}
Taking the limit $t \rightarrow \infty$ on both sides of the above inequality while realizing that $\lim_{t \rightarrow \infty}V(e(t)) = 0$ due to the stability of error dynamics and $V(e_0) = 0$ due to zero initial error, one can obtain the following result
\begin{align}
	&\int_{0}^{\infty}\hspace{-0.05cm}\norm{\m \epsilon(\tau)}^2_2 d\tau  \hspace{-0.05cm}<\hspace{-0.11cm}  \int_{0}^{\infty}\hspace{-0.05cm}\gamma \norm{\m w(\tau)}^2_2 d\tau\hspace{-0.05cm}\Leftrightarrow \hspace{-0.05cm}\norm{\m \epsilon(t)}^2_{L2} \hspace{-0.05cm}<\hspace{-0.05cm}\gamma \norm{\m w(t)}^2_{L2}.\nonumber 
\end{align}
Now, from \eqref{eq:robust_observer_proof_1}, the condition $\dot{V}(\m e) + \m {\epsilon}^\top \m {\epsilon} - \gamma \m w^\top\m w < 0$ can be shown to be equivalent to the following matrix inequality
\begin{align}
	\bmat{\m A^\top\hspace{-0.05cm}\m P + \m P^\top\hspace{-0.05cm}\m A - \m C^{\top}\hspace{-0.05cm}\m L^{\top}\hspace{-0.05cm}\m P -\m P^\top\hspace{-0.05cm}\m L \m C + \m \Gamma^\top\m \Gamma & * \\ \m B_w^\top \m P + - \m D_w^\top \m L^\top\hspace{-0.05cm}\m P & -\gamma \m I} \prec 0.\nonumber
\end{align}
Finally, by constructing $\m P$ such that $\m P := \m X \m E + \m E^{\perp\top}\m Y$ for some $\m X$ and $\m Y$ where ${\m E}^{\perp}$ being an orthogonal complement of $\m E$ \cite{xu2006robust} and defining a new matrix $\m W$ as $\m W := \m L^{\top}\m P$, the LMI \eqref{eq:H_infinity} is established. This concludes the proof.
\newqed
\vspace{-0.1cm}

\vspace{-0.2cm}
\section{Proof of Theorem \ref{thm:regular_aug}}\label{appdx:F}
Firstly, observe that for the pair $(\m E_{\xi},\m A_{\xi})$ we have
\begin{align}
	\det\left(s\m E_{\xi}-\m A_{\xi}\right) = \det\left(\bmat{s\m E-\m A & -\m B_{\nu}\\ \m O & s\m I}\right).\label{eq:regular_aug_proof_1}
\end{align}
Since it is assumed that $\m A_4$ in DAE \eqref{eq:DAE_base} is a full rank matrix, then due to Theorem \ref{thm:regular}, we know that the pair $(\m E,\m A)$ is regular and of index one. Thus, without loss of generality, there exists $s\in\mbb{C}$ where $s\neq 0$ such that $s\m E-\m A$ is nonsingular. As such, also from the right-hand side of \eqref{eq:regular_aug_proof_1}, it can be shown that 
%\begin{align*}
%\det\hspace{-0.05cm}\left(s\m E_{\xi}-\m A_{\xi}\right) = \det(s\m E-\mA)\hspace{-0.025cm}\det(s\m I) = s^{\nu}\hspace{-0.025cm}\det(s\m E-\mA),
%\end{align*}
%which leads to
\begin{align*}
	\deg \left(\det\hspace{-0.05cm}\left(s\m E_{\xi}-\m A_{\xi}\right)\right) &= \deg\left(\det(s\m E-\mA)\hspace{-0.025cm}\det(s\m I)\right)\\
	&= \deg \left(s^{\nu}\hspace{-0.025cm}\det(s\m E-\mA)\right) \\&= \nu + \deg \left(\det(s\m E-\mA)\right) \\ &= \nu + \rank(\m E) = \rank(\m E_{\xi}),
\end{align*}
hence suggesting that the pair $(\m E_{\xi},\m A_{\xi})$ is impulse-free, which further implies that $(\m E_{\xi},\m A_{\xi})$ is also of index one and regular  \cite{Dai1989}. Secondly, based on the detectability condition in \eqref{eq:detectability}, it is straightforward to show from \eqref{eq:DAE_base_aug} that
\begin{align}
	\rank\left(\bmat{s \m E_{\xi} -\m A_{\xi} \\ \m C_{\xi}}\right) &= \rank\left(\bmat{s \m E -\m A & -\m B_{\nu} \\ \m O & s\m I \\ \m C & \m O}\right)\nonumber \\
	&= \rank\left(\bmat{s \m E -\m A\\ \m C}\right) + \nu \nonumber \\ &= n+ \nu = \varsigma.\label{eq:detectability_aug_proof}
\end{align}
Thirdly, it also follows from I-observability condition \eqref{eq:I-observability} and the augmented linear DAE \eqref{eq:DAE_base_aug} that
\begin{align}
	\rank \left(\bmat{\m E_{\xi} & \m A_{\xi} \\ \m O & \m E_{\xi} \\ \m O & \m C_{\xi}}\right) 
	&= \rank \left(\bmat{\m E & \m O & \m A & \m B_{\nu} \\
		\m O & \m I & \m O & \m O \\ \m O & \m O & \m E &\m O \\
		\m O & \m O & \m O & \m I \\ \m O & \m O &\m C & \m O}\right)\nonumber \\
	&= \rank \left(\bmat{\m E & \m A\\ \m O & \m E \\ \m O & \m C}\right) + 2\nu\nonumber \\
	&= n + \rank(\m E) +  2\nu \nonumber \\ &= \varsigma + \rank(\m E_{\xi}).\label{eq:I-observability_aug_proof}
	%n + \rank(\m E). \label{eq:I-observability_aug}
\end{align}
Therefore, from \eqref{eq:detectability_aug_proof} and \eqref{eq:I-observability_aug_proof}, 
detectability and I-observability of DAE \eqref{eq:DAE_base_aug} imply the detectability and I-observability of power networks DAE \eqref{eq:DAE_base} and vice versa. This ends the proof.
\newqed

\end{document}